\documentclass[apj]{emulateapj}
\usepackage[backref,breaklinks,colorlinks,citecolor=blue]{hyperref}
\usepackage{lipsum}
\usepackage{float}
\usepackage{color}
\usepackage[section]{placeins}
\usepackage{epstopdf}

\newcommand{\vvc}{\Omega/\Omega_{\rm{crit,ZAMS}}}
\newcommand{\msol}{M_{\odot}}
\newcommand{\vsini}{V\sin i}

\shortauthors{GOSSAGE ET AL.}

\begin{document}

\title{Combined Effects of Rotation and Age Spreads on 
       Extended Main Sequence Turn Offs}
 
\author{Seth Gossage\altaffilmark{1}, 
        Charlie Conroy\altaffilmark{1},
        Aaron Dotter\altaffilmark{1},
        Ivan Cabrera-Ziri\altaffilmark{1,*},
        Andrew E. Dolphin\altaffilmark{2},
        Nate Bastian\altaffilmark{6},
        Julianne J. Dalcanton\altaffilmark{3},
        Paul Goudfrooij\altaffilmark{5},
        L. Clifton Johnson\altaffilmark{4},
        Benjamin F. Williams\altaffilmark{3},
        Philip Rosenfield\altaffilmark{1,7},
        Jason Kalirai\altaffilmark{5,8},
        Morgan Fouesneau\altaffilmark{9}
       }

\altaffiltext{1}{Harvard-Smithsonian Center for Astrophysics,
                 60 Garden Street, Cambridge, MA 02138, USA}

\altaffiltext{2}{Raytheon,
                 1151 E. Hermans Road, Tuscon, AZ 85706, USA}

\altaffiltext{3}{Department of Astronomy, University of 
                 Washington, Box 351580, Seattle, WA 98195, USA}

\altaffiltext{4}{Department of Physics and Astronomy, 
                 Northwestern University, 2145 Sheridan Road, 
                 Evanston, IL 60208, USA }

\altaffiltext{5}{Space Telescope Science Institute, 3700 
                 San Martin Drive, Baltimore, MD 21218, USA}

\altaffiltext{6}{Astrophysics Research Institute, Liverpool 
                 John Moores University, 146 Brownlow Hill, 
                 Liverpool L3 5RF, UK}

\altaffiltext{7}{Eureka Scientific, Inc., 2452 Delmer Street, 
                 Oakland CA 94602, USA}

\altaffiltext{8}{Johns Hopkins University Applied Physics 
                 Laboratory, 11101 Johns Hopkins Road, Laurel, 
                 MD 20723, USA; Jason.Kalirai@jhuapl.edu}

\altaffiltext{9}{MPIA Heidelberg, Heidelberg, Germany}

\altaffiltext{*}{Hubble Fellow}

%\slugcomment{Submitted to ApJ}
\slugcomment{Accepted to ApJ}

\begin{abstract}

The extended main sequence turn offs (eMSTOs) of 
several young to intermediate age clusters are 
examined in the Magellanic Clouds and the Milky 
Way. We explore the effects of extended star 
formation (eSF) and a range of stellar rotation 
rates on the behavior of the color-magnitude 
diagram (CMD), paying particular attention to the MSTO. 
We create synthetic stellar populations based on 
MESA stellar models to simulate observed \textit{Hubble 
Space Telescope} and \textit{Gaia} star cluster data. We 
model the effect of rotation as a non-parametric 
distribution, allowing for maximum flexibility. 
In our models the slow rotators comprise the 
blueward, and fast rotators the redward portion of 
the eMSTO. We simulate data under three scenarios: 
non-rotating eSF, a range of rotation rates with a 
single age, and a combination of age and rotation 
effects. We find that two of the five clusters (the 
youngest and oldest) favor an age spread, but these 
also achieve the overall worst fits of all clusters. 
The other three clusters show comparable statistical 
evidence between rotation and an age spread. In all 
five cases, a rotation rate distribution alone is 
capable of qualitatively matching the observed eMSTO 
structure. In future work, we aim to compare our 
predicted $\vsini$ with observations in order to 
better constrain the physics related to stellar 
rotation. 

\end{abstract}

\section{Introduction}
\label{sec:intro}

In the past decade, \textit{Hubble Space Telescope} 
(\textit{HST}) observations helped reveal the eMSTO 
phenomenon, now considered a 
ubiquitous feature of young star clusters in the 
Magellanic Clouds (e.g., \citealt{ADMBNP2007, 
ADM2008, KG2008, APM2009, PG2009}), i.e., those 
with ages less than 2 Gyr. However, the origin of the 
eMSTO has remained a mystery. Recently 
revealed to be common in clusters of the 
Milky Way by \textit{Gaia} (e.g., 
\citealt{AFM2018a, GC2018, BN2018}) as well, 
suggesting that the eMSTO is a natural property of 
young star clusters, rather than a peculiarity of 
the Magellanic Clouds. 

The eMSTO appears in a 
star cluster's CMD as a 
broadened MSTO, as if the cluster hosts multiple 
single age MSTOs. At face value, this suggests that 
such clusters contain stars born over some period of 
eSF in the past (e.g., \citealt{RuSt2010, 
RuSt2011, PG2011b, PG2011a, CCDS2011, SK2011, 
ADM2013}), perhaps lasting a few hundred Myr. 
Thus, first instincts were to view the eMSTO as 
a consequence of eSF, which was surprising under 
conventional beliefs that regard young clusters as 
simple stellar populations (SSPs) with coeval stars. 
Alternate theories have arisen since then, but a 
definitive consensus on the cause of the eMSTO has 
not been met. Ultimately, uncovering the physical 
process(es) behind this phenomenon could open new 
chapters in stellar evolution and cluster formation.
 
Much effort to reveal the true origin of the eMSTO 
has focused on searching for observational evidence 
of eSF. Such evidence could elucidate the 
contemporaneous puzzle of ``multiple populations'' 
(MPs) observed in globular clusters (GCs) older 
than 2 Gyr, as reviewed by \cite{NBCL2018}. 
Theoretically, eSF can reproduce observed eMSTO 
morphologies (e.g., \citealt{CL2017}), but it is 
challenged by a lack of empirical support, despite 
numerous observational campaigns. For example, 
\cite{ICZ2016} found no evidence for eSF events 
when modeling the spectral properties of the 
young massive cluster W3 in NGC 7252, despite this 
cluster's young age and exceptionally large escape 
velocity (possibly conducive to entrapment of 
stellar ejecta and subsequent eSF). Relatedly, 
\cite{APNB2016} and \cite{GC2018} found that many 
young LMC and Galactic clusters possess eMSTOs as 
well, in spite of their relatively low masses (e.g., 
around 2400 $\msol$ for NGC 2818, \citealt{BN2018}); 
presenting additional evidence that eMSTOs exist 
independent of the cluster's potential for gas 
retention. Observations have not revealed how 
these young clusters might retain enough gas to 
experience a (up to) several hundred Myr episode 
of star formation required to explain the largest 
eMSTOs. This is especially so, given that much more massive 
clusters like W3 (roughly 600 Myr old and 
$1.13\times10^{8}\msol$, \citealt{ICZ2016}, and 
see also \citealt{CZ2014}) appear to lack 
evidence of eSF.

An alternative explanation has 
developed alongside the eSF theory. \cite{BdM2009} 
made a case for stellar rotation as being the 
cause of the eMSTO. Rotation can grant stars a 
greater core fuel supply, thereby extending main 
sequence (MS) lifetimes (see e.g., 
\citealt{MeMa2000, MaMe2010, LG2011, SE2012, 
JC2016}). Rotation also causes gravity 
darkening, i.e., structural deformations that alter 
the apparent magnitudes and colors of rotating 
stars in a viewing angle dependent manner, see e.g., 
\cite{HvZ1924, LBL1967, ELR2011}. Combined, 
these effects allow coeval rotating stars to 
take on a range of colors and magnitudes that 
can mimic an age spread (e.g., \citealt{BH2015c, 
SG2018}) and create a broad MSTO. Additional 
observational evidence has accumulated for 
stellar rotation within the eMSTO (e.g. 
\citealt{BdM2009, ZL2012, WY2013, NF2015, 
BH2015c, BN2016, NF2016, BN2018, FDAn2018}, and 
\citealt{CG2019}). Furthermore, 
\cite{JoC2019} have recently shown, using non-rotating 
stellar models calibrated on \textit{Kepler} astroseismic 
data, that convective core overshooting can produce an eMSTO 
similar to observations. Thus, rotation and convective 
mixing are observed stellar processes at play in these 
clusters that may contribute to the eMSTO; we limit our 
study to the exploration of stellar rotation and age 
spreads as a matter of simplicity, but acknowledge that 
variable convective mixing efficiencies may also play a 
role.
 
A central finding in eMSTO studies has been that as 
clusters age, the eMSTO width increases (i.e., the 
$\Delta$(Age)-age trend highlighted by 
\citealt{NF2015}). This behavior may 
suggest that eMSTO width is related to stellar 
evolution, possibly the evolution of rotation. This 
interpretation says that the eMSTO is not due to a 
range of ages, but rather coeval stars with a 
distribution of rotation rates. If the 
$\Delta$(Age)-age trend is due to eSF, it would 
require that SF took longer in the past. This 
explanation is not impossible, but it is difficult 
to find a natural reason as to why the length of SF 
has steadily decreased over time. Furthermore, 
\cite{GC2018} have found a similar $\Delta$(Age)-age 
trend for young Milky Way clusters, suggesting that 
the trend is not specific to the environments of 
the LMC/SMC. Many lines of evidence now point 
towards stellar rotation at least playing some 
role in causing the eMSTO. This is often 
acknowledged, but it is still argued that some 
degree of eSF may be present in these clusters.

% cite milone for binary frac.
Hiding underneath all of this could 
be additional effects due to stellar 
binary interactions. For instance, binary 
mergers could replenish stars nearing 
terminal age MS (TAMS), or otherwise alter 
their evolutionary trajectory. Relatively 
fewer studies have incorporated models 
that account for binary evolution, but 
work has been done by e.g., \cite{WY2011} 
and \cite{ZL2012, ZL2016}. As high mass 
stars can still exist in young clusters, 
binary effects may be especially important 
there (e.g., \citealt{WY2018} and 
suggested by \citealt{ERB2019}). Binary 
fractions are still expected to be 
significant ($\sim30$\% in many studies) in 
intermediate age clusters (1-2 Gyr), 
so some degree of binary interaction could 
be expected, but may be minimal. We 
acknowledge that binary interaction has a 
potentially strong effects, but do not model 
it in this work. 

Along these lines, \cite{FDAn2017} have suggested 
that braking due to binary interaction may 
contribute to the rotation rate evolution of 
cluster stars and result in populations of fast 
and slow rotators. Therefore, this mechanism could 
be part of the physical explanation not only for 
eMSTO spreads, but also for the observed split 
MS in younger star clusters. The split MS shows 
up in a CMD as a blue/redward bifurcation of the 
main sequence. It is now widely thought to be 
due to a bimodal distribution of stellar 
rotation rates (see e.g., \cite{APM2018} for 
examples). \cite{AFM2018b} found first spectroscopic 
evidence for the split MS hosting slow and fast 
rotators in the redward and blueward portions of 
the split MS, respectively, for the young 
globular cluster NGC 1818. This feature does not 
appear to be due to metallicity variations, 
nor photometric errors, and has a numerically dominant 
population in the redward MS (around 60-70\% of the 
split MS stars), which are the proposed fast rotators 
(\citealt{APM2016,APM2017}). Star clusters younger 
than about 600 Myr (\citealt{APM2018}) appear to 
host a split MS, in addition to their eMSTO. In our 
cluster sample, NGC 1866 is one such cluster. Our 
analysis of its split MS is highlighted in later 
sections. Whether or not the binary braking 
scenario of \cite{FDAn2017} causes this is 
uncertain at this point, rotation distributions 
in the split MS may reflect the initial distribution 
that stars are born with instead.

Until now, a quantitative assessment of the 
CMD morphology with flexible stellar models 
has been missing, and we aim to provide this 
here. Previous studies have variously used 
stellar models that are non-rotating, e.g., 
PARSEC (\citealt{AB2012}), with rotation often 
modeled according to either GENEC 
(\citealt{SE2012}), or MESA (\citealt{BP2013}). 
PARSEC has traditionally been a non-rotating model set 
(though see \citealt{GCosta2019}), while GENEC 
provides a dense grid of stellar rotation 
rates, but is limited in stellar mass to 
$1.7\ \msol$ and above, and a relatively coarse 
metallicity sampling. GENEC has been the 
workhorse model set in these studies for 
including the effects of stellar rotation, 
and has done much to form our current 
understanding. We have extended MIST to now 
include a dense grid of rotation rates. 
Meanwhile, in comparison to GENEC, our new 
MIST model set also includes masses ranging 
down to $0.1\ \msol$, and finer metallicity 
sampling. These new rotating stellar models 
can now investigate the effects of stellar 
rotation in older clusters than previous 
studies. We form synthetic stellar populations 
that can span a range of ages, metallicities, 
and stellar rotation rates. In our analysis, we 
opt for a non-parametric rotation rate 
distribution whose predictions are presented 
in our results.

We explore the two main scenarios proposed to 
explain the eMSTO: eSF and stellar rotation. 
To this end, we build synthetic stellar 
populations according to three scenarios: 1) 
populations that experienced eSF, or 2) 
experienced no eSF but could have stars at 
a range of rotational velocities, and 3) a 
cluster that experienced both eSF and has 
stars at a range of rotational velocities. 
We fit these synthetic populations to 
observations of the clusters NGC 1866, 1831, 
2818, 2249, and 2203 and determine the 
best fit model populations, thereby deriving 
cluster properties like age, age spread, and 
the presence of rotation distributions. In 
Section \ref{sec:data}, we present the 
sources for our data and give brief 
descriptions of each cluster. Section 
\ref{sec:methods} provides details of our 
models, lays out our fitting procedure, and 
presents mock tests of the methods. Section 
\ref{sec:results} gives our results, 
sequentially for each scenario along with 
brief commentary. Finally, our conclusions 
are summarized in Section \ref{sec:conclusions}, 
along with some discussion of caveats and suggestions 
for future work. The photometric zero point is Vega 
for all magnitudes shown.

\begin{deluxetable}{cccccc}[!t]
\tablecaption{Adopted Cluster Parameters}
\tablehead{ \colhead{Cluster} & \colhead{$\mu$\tablenotemark{1}} 
          & \colhead{Age [Gyr]} & \colhead{[Fe/H]\tablenotemark{2}}
          & \colhead{$A_{\rm{V}}$} & \colhead{Binary \%}}

\startdata
  NGC 2203 & 18.37 & 1.55 & -0.30 & 0.16 & 18 \\
  NGC 2249 & 18.20 & 1.0 & -0.46 & 0.07 & 30 \\
  NGC 2818 & 12.76 & 0.7 & 0.0 & 0.90 & 29 \\
  NGC 1831 & 18.35 & 0.7 & -0.25 & 0.14 & 20 \\
  NGC 1866 & 18.31 & 0.2 & -0.36 & 0.34 & 25 \\
  \enddata
  \vspace{0.1cm}
  \tablecomments{All parameters listed are fixed in our fits, 
                 except the age. Age initialized at the listed 
                 value but is allowed to vary. See the text for 
                 the literature sources to these values.}
  \tablenotetext{1}{Distance modulus}
  \tablenotetext{2}{Based on \cite{MA2009} protosolar abundances}

\label{t:cparams}
\end{deluxetable}

\section{Data}
\label{sec:data}

\begin{figure*}[!ht]
  \center
    \includegraphics[width=\textwidth]{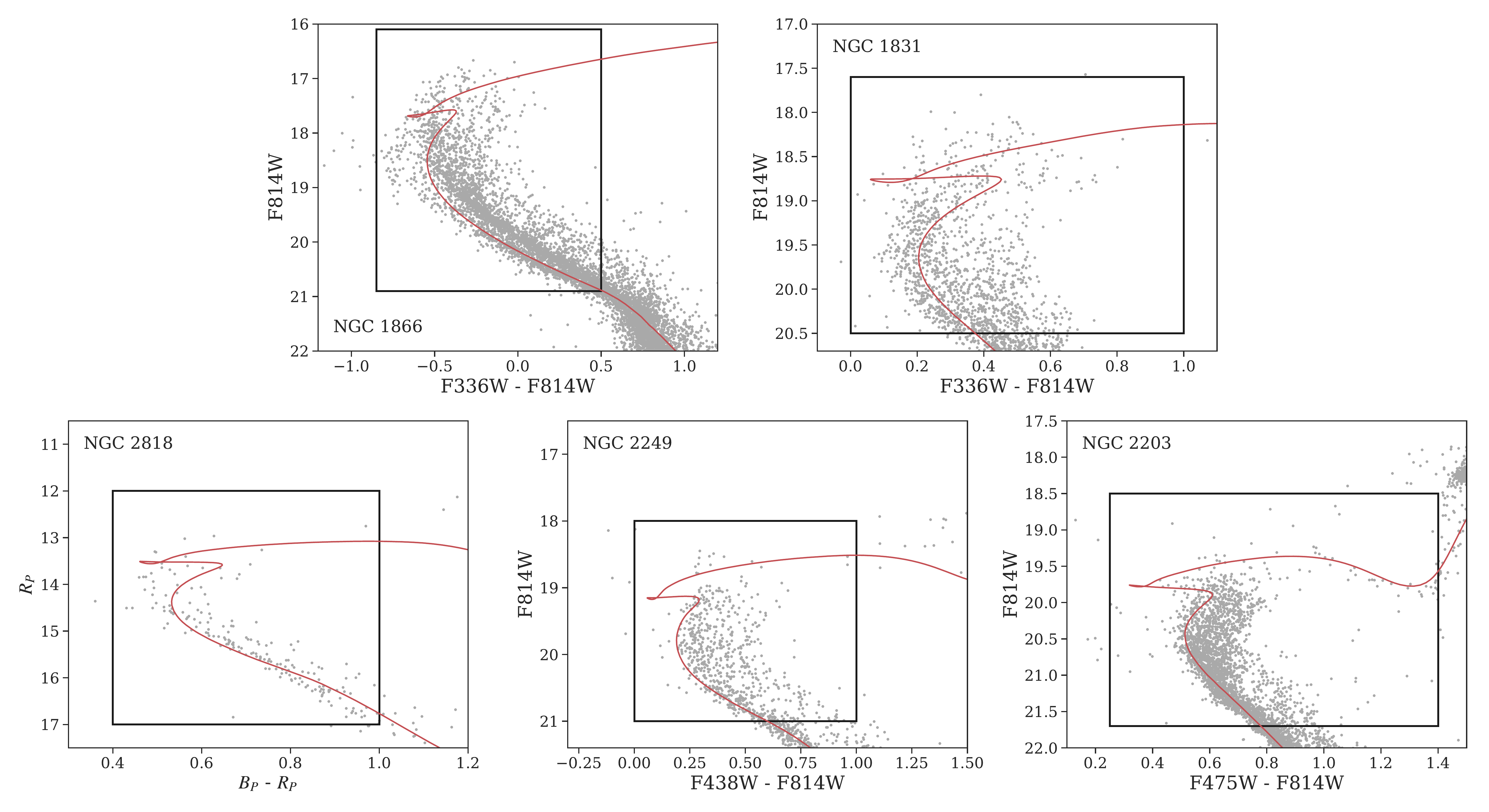}
    %\vspace{0.1cm}
    \caption{CMDs of our five chosen star clusters, 
             in order of descending age: NGC 1866, 1831, 2818, 2249, 2203. 
             Black rectangles show the CMD area used for fitting 
             models to data. Red (non-rotating, MIST) isochrones 
             represent SSPs according to our adopted cluster parameters, 
             listed in the text. The red clump is ignored in our fits, for 
             reasons explained in the text.}
\label{fig:photometry}
\end{figure*}

Our data comes from \textit{HST} 
observations and the recent \textit{Gaia} 
DR2. Our target clusters are NGC 2203, 2249, 
1831, and 1866, located in the LMC, and 
NGC 2818, located in the Milky Way. We chose 
this set of clusters because they cover a 
good portion of the age range where eMSTOs 
are observed, i.e., younger than about 1.5 Gyr. 
As discussed in Section \ref{sec:methods}, our 
models are limited to $<5\ \msol$, and so we 
cannot model clusters younger than about 200 Myr 
currently. For each cluster, we list the values of 
mean log age, distance modulus, binary fraction, 
$A_{\rm{V}}$, and metallicity [Fe/H]; the 
mean log age listed here serves to inform 
the initial position of our parameter 
search in age during our fits, while we 
keep all other parameters fixed. The adopted 
parameters are collected in Table \ref{t:cparams}. 
CMDs produced by the data are shown in Figure 
\ref{fig:photometry}, where black, dashed 
boxes show the regions we fit models to 
(focusing on the MSTO, ignoring the red 
clump, or RC). It is known that some 
models can have trouble simultaneously 
fitting the RC and MSTO (e.g., 
see \citealt{BN2018}) and this is the reason 
we exclude it in our fits. Red isochrones 
are also plotted, showing the SSP expected 
by the adopted cluster parameters, which 
are listed in the following subsections.

\subsection{NGC 2203}
\label{ssec:ngc2203_data}

The intermediate age cluster NGC 2203, 
located in the LMC is the oldest of our 
target clusters according to the literature, 
at about 1.55 Gyr (\citealt{PG2014, 
PR2017}). Photometry for this cluster 
is the same as that used in 
\cite{PR2017}, where the data 
reduction process is also described. 
In brief, this data is part of a 
larger set that was re-reduced from 
two \textit{HST} programs: GO-9891 
(PI: Gilmore) and GO-12257 (PI: 
Girardi). So, this photometry comes 
from archival ACS and WFC3 data, 
re-reduced with the University of 
Washington data reduction 
pipeline, designed to reduce the 
\textit{HST} treasury programs ANGST 
(\citealt{JJD2009}) and PHAT 
(\citealt{JJD2012}); see 
\cite{BFW2014} for further details. 
\texttt{ASteCA} (\citealt{PVP2015}) was 
used to determine the 
cluster center, found via the maximum 
spatial density using a 2D Gaussian kernel 
density estimator. The cluster radius 
corresponds to where the radial density 
profile becomes indistinguishable 
from the background stellar density. 
\texttt{ASteCA} was also used to determine 
contamination, utilizing a non-parametric 
Bayesian decontamination algorithm based 
on the method of \cite{CCA1990}. The final 
membership was limited to stars within the 
cluster radius with $>70\%$ membership 
probability. 

We adopt the cluster parameters 
cited in \cite{PG2014}, i.e., their 
age, [Fe/H], $A_{V}$, and distance 
estimated via best-fit isochrones 
from the Padova group (\citealt{PM2008}). 
Respectively, these are values of 
1.55 Gyr, -0.30, 0.16 dex, and 18.37. 
We also adopt the binary fraction of 
0.18 used by \cite{PG2014}. Taking into 
account that slight differences in fits 
to CMD features can arise due to different 
input physics between our MIST and those 
Padova models. Although, see that in 
Figure \ref{fig:photometry} that these 
parameters still provide a reasonable fit 
with our models.

%\newpage
\subsection{NGC 2249}
\label{ssec:ngc2249_data}

On the boundary of the intermediate 
age regime, NGC 2249 is a star
cluster also located in the LMC. 
Membership for this cluster was 
determined in an identical way to 
what was described for NGC 2203.
Isochrone analysis in \cite{MC2014} 
has estimated the cluster to be 
about 1 Gyr old. Data reduction for 
this cluster is the same as 
described for NGC 2203, with further 
details in \cite{PR2017}. 
In this work, we adopt the 
same mean age, [Fe/H], $A_{\rm{V}}$, 
and distance modulus for NGC 2249 as 
cited by \cite{MC2014}; namely, 1 
Gyr, -0.46, 0.07 dex, and 18.2 mag, 
respectively. We adopt the binary 
fraction of 0.30 cited by 
\cite{MC2014}.

\subsection{NGC 2818}
\label{ssec:ngc2818_data}

NGC 2818 enters the regime of 
young cluster ages for our target 
clusters, at about 700 Myr, and it 
is located in the Milky Way. Our 
NGC 2818 data is taken from 
the publicly available data 
gathered by \cite{BN2018}, 
originating from \textit{Gaia} 
DR2 (\citealt{GAIA2016, 
GAIA2018}). Details of the member 
selection for this cluster are 
given in \cite{BN2018}, but briefly, 
members were selected via proper 
motion and parallax cuts. NGC 2818 
is one example of a number of young 
Galactic clusters with eMSTOs that 
have been revealed with \textit{Gaia} 
DR2 data in the last year or so, 
thanks to \textit{Gaia}'s enhanced 
photometric precision. The adopted 
cluster parameters are based on 
values used in \cite{BN2018}: with an 
age of 700 Myr, a solar [Fe/H] of 0.0, 
$A_{\rm{V}} = 0.90$, distance modulus 
of 12.76. We adopt a binary 
fraction of 0.29 from \cite{GC2018}.

\subsection{NGC 1831}
\label{ssec:ngc1831_data}

NGC 1831 is also an approximately 
700 Myr old cluster as well, 
located in the LMC. For this 
cluster, we carried out PSF 
photometry on the flat-field 
corrected, and bias-subtracted 
\textit{HST} `flc' images (Program ID: 
GO-14688) using the WFC3 module of 
\texttt{DOLPHOT}, a modified 
version of \texttt{HSTphot} 
(\citealt{AD2000}) and following the 
procedure described in \cite{EB2009}. 
For our analysis, and for the CMD 
shown in Figure \ref{fig:photometry}, 
we chose stars within a half-light 
radius (from \citealt{McLvdM2005}), 
$r_h=33.85''$ of the cluster's center. 
We have adopted the parameters based 
on those used by \cite{PG2018} for 
NGC 1831, determined from PARSEC 
isochrones. Thus, an age of 700 Myr, 
[Fe/H] of -0.25, $A_{\rm{V}}$ of 0.14, 
distance modulus of 18.35, and a 
binary fraction of 0.20.

\subsection{NGC 1866}
\label{ssec:ngc1866_data}

NGC 1866 is the youngest cluster 
in this study, at about 200 Myr old, 
and it is located in the LMC. 
This photometry was obtained in the 
same manner as described for NGC 
1831 (except from \textit{HST} Program 
ID: GO-14204). Here we have taken stars 
within the half-light radius 
$r_h=42.9''$ of the cluster's 
center, based on \cite{McLvdM2005}. 
Our adopted cluster parameters 
for NGC 1866 are adopted from 
\cite{APM2017} for the metallicity 
and binary fraction. These 
parameters are an age of 200 Myr, 
[Fe/H]=-0.36, $A_{\rm{V}}=0.34$, a 
distance modulus of 18.31, and a 
binary fraction of 0.25.

\section{Methodology}
\label{sec:methods}

Major aspects of our models are identical 
to those described in \cite{SG2018}, but 
topical details are recapitulated, and 
additions are noted here. The greatest 
difference in our current models and 
those described in the aforementioned 
paper is the inclusion of higher 
rotation rates. Our mass range has 
changed and is from $0.1-5\ \msol$, with 
metallicities from [Fe/H] = $-0.60$ to 
$0.45$ in $0.15$ dex steps. This mass 
range extends lower than what is 
available in SYCLIST (\citealt{CG2014}, 
a stellar population synthesis tool, and 
stellar isochrone and track database). The 
SYCLIST isochrones and tracks are (similar 
to how MIST models are based on MESA) based 
on the GENEC stellar evolution code whose 
lowest stellar mass is $1.7\ \msol$. 
SYCLIST has been the primary model set 
used to study stellar rotation in 
these clusters so far. As our new models 
extend to 0.1 $\msol$, we can model 
older clusters than SYCLIST has traditionally 
been able to, however, with 
the caveat that magnetic braking is crudely 
accounted for in these lower mass stars (more 
on this in Section \ref{ssec:mesamods}). We 
do not evolve binary systems, or investigate 
the effects of mass transfer, nor tidal braking 
in our model set. Our models also allow for 
a finer metallcity range than GENEC does 
publicly, which offers Z = 0.014, 0.006, 
and 0.002 (roughly [Fe/H] = 0.0, -0.35, and 
-0.85).

\begin{figure*}[!th]
  \center
    \includegraphics[width=0.95\linewidth]{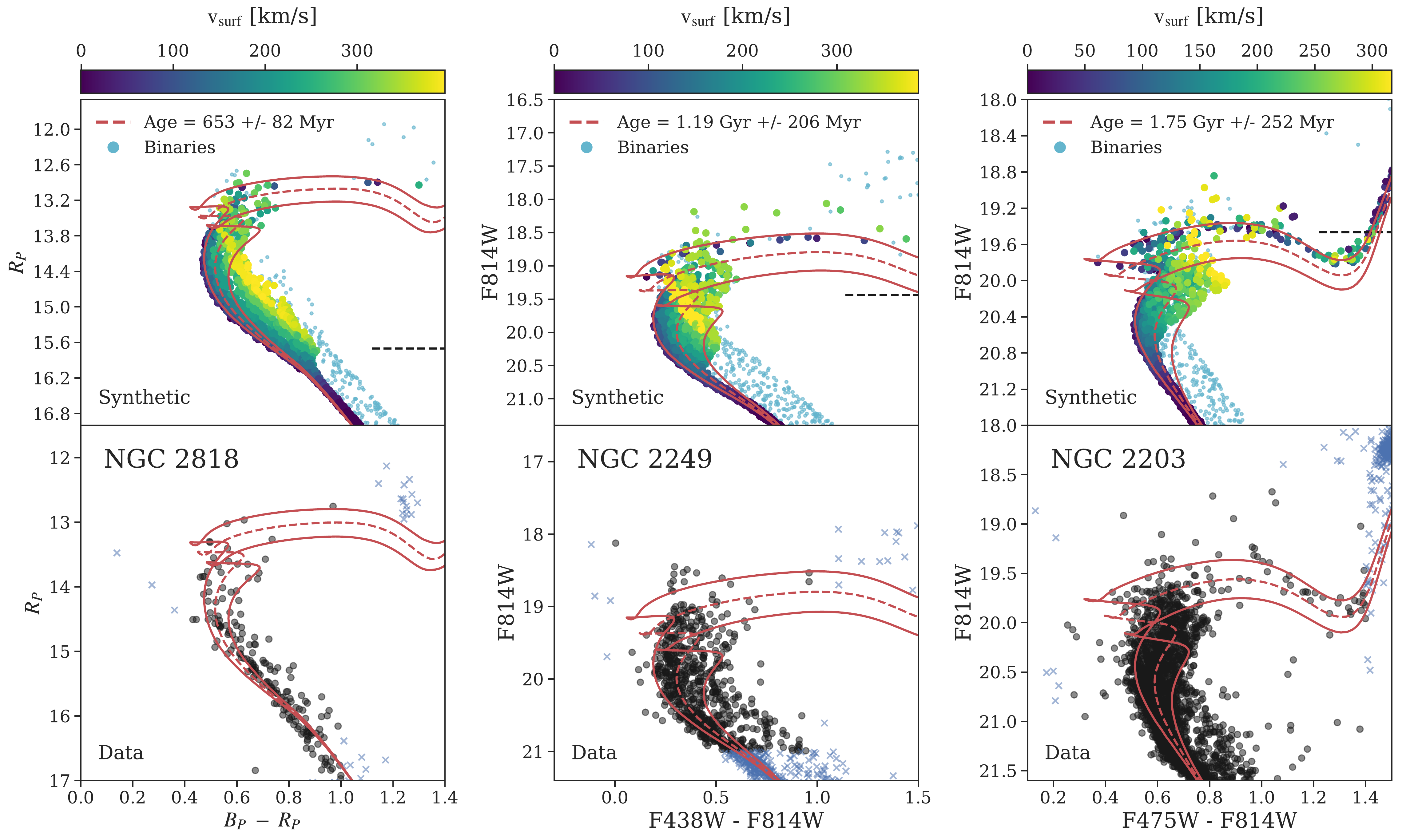}
    %\vspace{0.1cm}
    \caption{\textbf{Top row}: synthetic clusters, with 
    each point color coded by the stellar model's average 
    surface velocity. These points were generated at a single 
    age of 0.7, 1, and 1.5 Gyr, from left to right. Binaries 
    are shown as cyan points. Solid red lines mark the 
    youngest and oldest age non-rotating MIST isochrones that 
    span the eMSTO width; rotation effects can mimic this 
    span. The red dashed lines show the mean age. Turn off 
    masses (left to right): 1.5 - 1.8 $\msol$, 1.60 - 2.15 
    $\msol$, 2.1 - 2.85 $\msol$, respectively, according to 
    our models. The black dashed line approximately marks 
    the magnitude below which our velocities are ramped down 
    as a proxy for magnetic braking. \textbf{Bottom row}: data 
    for NGC 2818, 2249, and 2203 as black dots, with the same 
    non-rotating isochrones from above overlaid for comparison. 
    Blue crosses are data excluded from the fits.}
\label{fig:fakephot}
\end{figure*}

\subsection{Stellar Population Models}
\label{ssec:mesamods}

The starting point for our models is 
the \texttt{MESA} stellar 
evolution code (\citealt{BP2011, BP2013, 
BP2015, BP2018, BP2019}), version 
\texttt{r7503}, which is a modular and 
open source 1D stellar evolution code. 
We closely followed the physics used 
for the MIST database (\citealt{JC2016}), 
adopting the protosolar abundances of 
\cite{MA2009} and using boundary 
conditions from ATLAS12, while SYNTHE is 
used for bolometric corrections 
(\citealt{RLK1970, RLK1993}). Our models 
are evolved to the end of core helium 
burning. Hereafter, we will refer to our 
models as ``MIST models'' or 
``MIST-based''.

The MIST models are set rotating at 
the zero age MS (ZAMS) with a given 
velocity denoted by the ratio of 
equatorial angular velocity 
$\Omega_{\rm{ZAMS}}$ at ZAMS, over the 
critical $\Omega_{\rm{c}}$. The 
critical velocity $\Omega_{\rm{c}}$ is 
a property intrinsic to the star that 
depends on its mass (see e.g., 
\citealt{AM2009}); it represents the 
limit where centrifugal force overcomes 
the star's gravity. 
The ratio $\vvc$ is equivalent to the 
linear velocity form $v/v_{\rm{c,ZAMS}}$ 
in the \texttt{MESA} formalism. 
Previously, our models were limited to 
$\vvc = 0.6$ in \cite{SG2018}; we 
now include models ranging up to 
$\vvc = 0.9$, in steps of $0.1$. The 
initial MIST models released 
by \cite{JC2016} only included 
non-rotating and $\vvc=0.4$ models.

MIST models rotation under the shellular 
approximation developed by \cite{KT1970}, 
with chemical and angular momentum 
transport described by the equations of 
diffusion equations of \cite{ES1978}. This 
diffusive formalism is also adopted in the 
stellar evolution codes STERN 
(\citealt{IB2011}) and the recent version of 
PARSEC (\citealt{GCosta2019}). The 
shellular approximation is standard in 1D 
stellar evolution codes. The treatment 
of angular momentum and chemical transport varies 
between codes. For instance, GENEC uses a 
diffusive-advective formalism, described in 
\cite{JPZ1992, MaZh1998, MaMe2000a}. The 
differences in these two formalisms have 
significant effects on the models, e.g., 
leading to different MS lifetime extensions 
(by up to 20\% or so) and color-magnitude 
variations (see \citealt{JC2016, SG2018} for 
examples). Our models possess a stronger convective 
mixing with weaker rotation mixing, whereas 
GENEC features the opposite. Consequently, 
our rotating models are primarily affected 
by the structural changes of gravity 
darkening when they rotate, they don't see 
a dramatic MS lifetime extension or 
luminosity enhancement from rotational 
mixing, as is seen in the GENEC models. 

Gravity darkening is handled by the 
equations of \cite{ELR2011} (recently 
adopted by \cite{BP2019} in MESA as well) 
in determining the (surface averaged) 
luminosity and temperature of a given 
stellar model at viewing angle $i$. The 
viewing angle corresponds to $i=90^{\circ}$ 
when viewing is equator-on, versus 
$0^{\circ}$ when viewing pole-on. Gravity 
darkening is the effect of centrifugal 
force reducing the surface gravity of a 
rotating star. This effect is stronger at 
the equator than at the poles, due to the 
greater centrifugal force at the equator. 
Thus, gravity darkening causes the 
equator of a rotating star to become 
cooler and dimmer than the poles, 
introducing a viewing angle dependence on 
the apparent magnitude and color of a 
rotating star. The effects can be 
substantial; examples 
for our models exist in \cite{SG2018}. 
The formalism that we adopt from 
\cite{ELR2011} is similar to the gravity 
darkening formalism used for the 
GENEC-based SYCLIST stellar population 
models.

Another important aspect of rotation, at 
least for masses $\lesssim1.8\ \msol$ 
is magnetic braking. Modeling this process 
is an active area of research (e.g., 
\citealt{CeGa2016, SA2017, CeGa2018, 
JF2019}). \cite{NF2015} predicted the 
$\Delta$(Age)-age trend should stop after 
magnetic braking becomes effective. This 
limit is expected to be reached by TO stars in 
clusters with ages older than about 1.5 
Gyr, depending on the metallicity. In 
their recent work, \cite{CG2019} used 
models developed with the STAREVOL code 
(\citealt{LA2016}), for masses between 1 
and 2 $\msol$, including a prescription 
for magnetic braking according to 
\cite{SPM2015}. This mass range is roughly 
where surface convection zones develop, 
leading to surface magnetic fields that can 
act on extended stellar material, 
braking the star. We take a crude approach to 
simulate this, in absence of a proper 
model of the effects of magnetic braking. 
Below $M = 1.3\ \msol$, models are forced to 
be non-rotating; from $M = 1.3-1.8\ \msol$ 
models have $\vvc$ scaled up to the full 
value. The SYCLIST models do not 
model magnetic braking, but also exclude 
stellar masses below $M = 1.7\ \msol$ 
(\citealt{CG2014}). The MSTOs of NGC 2203 
and 2249 have TO masses that are low enough 
for magnetic braking to become important, so our 
results for these clusters in particular 
will be affected by uncertainties due to 
magnetic braking.

From our MIST-based stellar models, we compute 
synthetic stellar populations, as in 
\cite{SG2018}, using the code MATCH 
\citep{AD2001}. Specifically, we use MATCH to 
compute Hess diagrams of CMDs, including 
unresolved binaries, at fixed values of Z, 
age, and $\vvc$. Photometric errors are 
simulated with MATCH via artificial star 
tests. The populations are created at 
distinct ages, covering log Age = 8.0 to 
9.5 (in 0.02 dex steps), each of which is 
also created at $\vvc=0.0,\ 0.1,...$, up to 
0.9. Our synthetic populations include the 
effect of gravity darkening via randomly 
drawn viewing angles for constituent stars. 
Stellar models are drawn according to a 
\cite{PK2001} initial mass function (IMF). 
We combine these synthetic populations 
(weighting them as described below 
in Section \ref{ssec:fitmethod}) to form a 
composite stellar population that may possess 
stars from a range of ages and rotation rates. 

Colored points in the top row of Figure 
\ref{fig:fakephot} shows several examples 
of these models. In this figure, models have 
fixed age pertaining to the representative 
cluster, from left to right: NGC 2818, 2249, 
and 2203. The colors map the full range 
of rotation rates (i.e., $\vvc=0.0$ 
to $0.9$), according to a flat distribution of 
$\vvc$, to clearly show the full effects of 
stellar rotation on the MSTO in these clusters. 
Red lines show non-rotating MIST-based isochrones, 
dashed at the mean age, and solid at $\pm\Delta$(age). 
This $\Delta$(Age) was chosen so that the isochrones 
roughly covered the full extent of the eMSTO. The 
bottom row of panels shows the same non-rotating 
isochrones overlaid on the data of NGC 2818, 2249, 
and 2203. Broadly, both stellar rotation and a range 
of ages can cover the extent of the eMSTO, but in 
different ways, hence the contention between the 
theories. 

Figure \ref{fig:fakephot} shows how increasing age 
reddens and decreases the luminosity of TO stars in a 
similar manner to the effects of gravity darkening. 
The effects of stellar rotation manifest in a confined 
region of CMD space at a given age, in comparison to 
an age spread which can modify the luminosity and 
temperatures of stars over as wide a space as is 
useful for covering the eMSTO. In this sense, the 
effects of an age spread have relatively more freedom 
than the effects of a rotation rate distribution. 
MATCH was used to generate the models displayed in the 
top row, but the code currently lacks the ability to 
perform a straightforward fit for rotation rate 
distributions. We opted for a separate fitting method 
outside of what is provided by MATCH (outlined below in 
Section \ref{ssec:fitmethod}) to handle this.

\begin{figure*}[!htb]
  \center
    \includegraphics[width=0.95\linewidth]{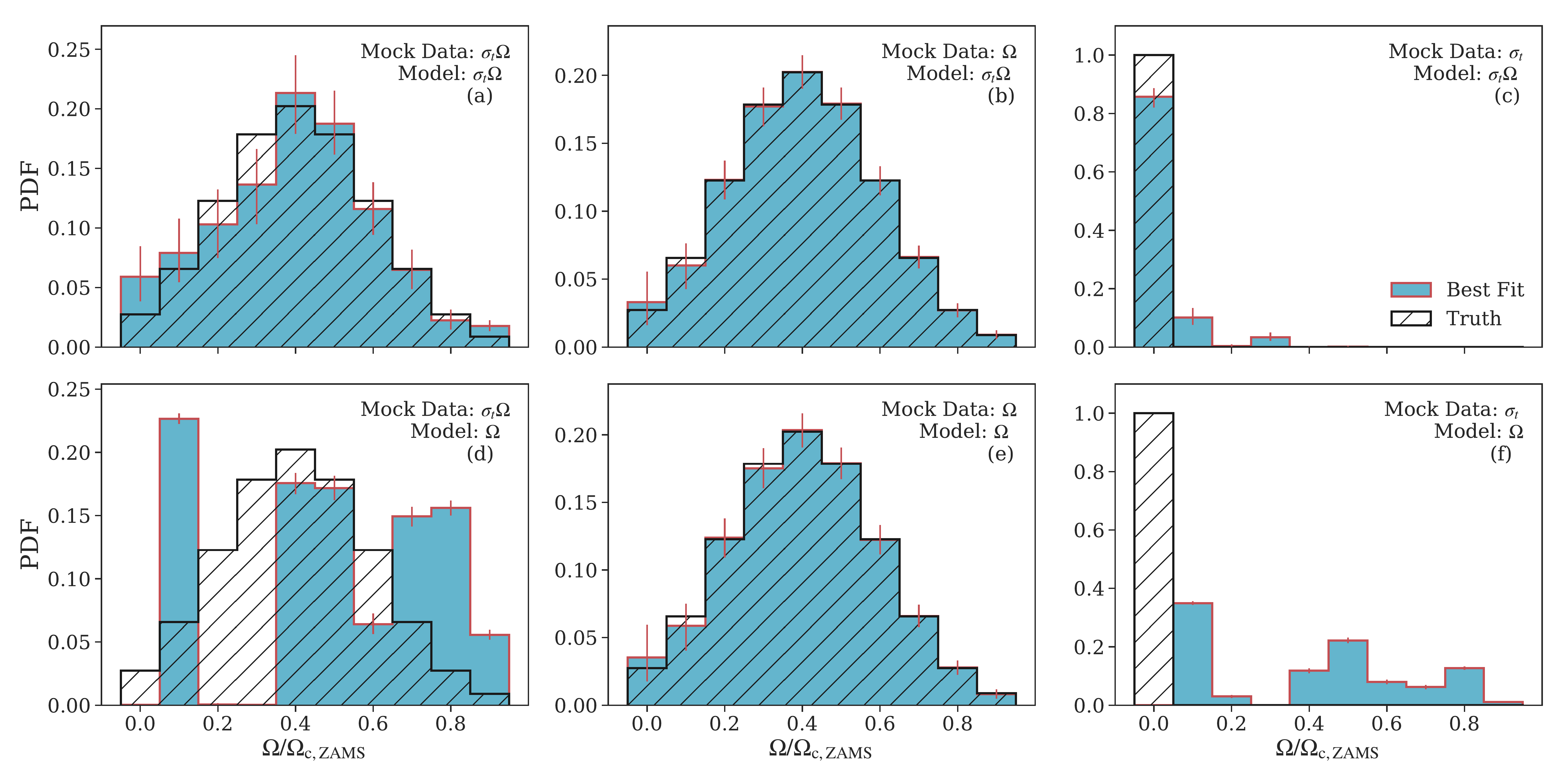}
    %\vspace{0.1cm}
    \caption{Recovered rotation rate distributions from 
             mock tests described in the text. Black hatched bins represent 
             the input ``true'' values, blue bins are best-fit distributions, with 
             error bars (84th, 16th percentiles) shown as red vertical lines. The 
             top row corresponds to cases where the $\sigma_t\Omega$-model was 
             tested against mock data created with the scenario written in the 
             annotations and described in the text. The bottom row shows the same, 
             but for the $\Omega$-model tests.}
\label{fig:mock_weights}
\end{figure*}

\begin{figure}[!htb]
  \center
    \includegraphics[width=0.95\linewidth]{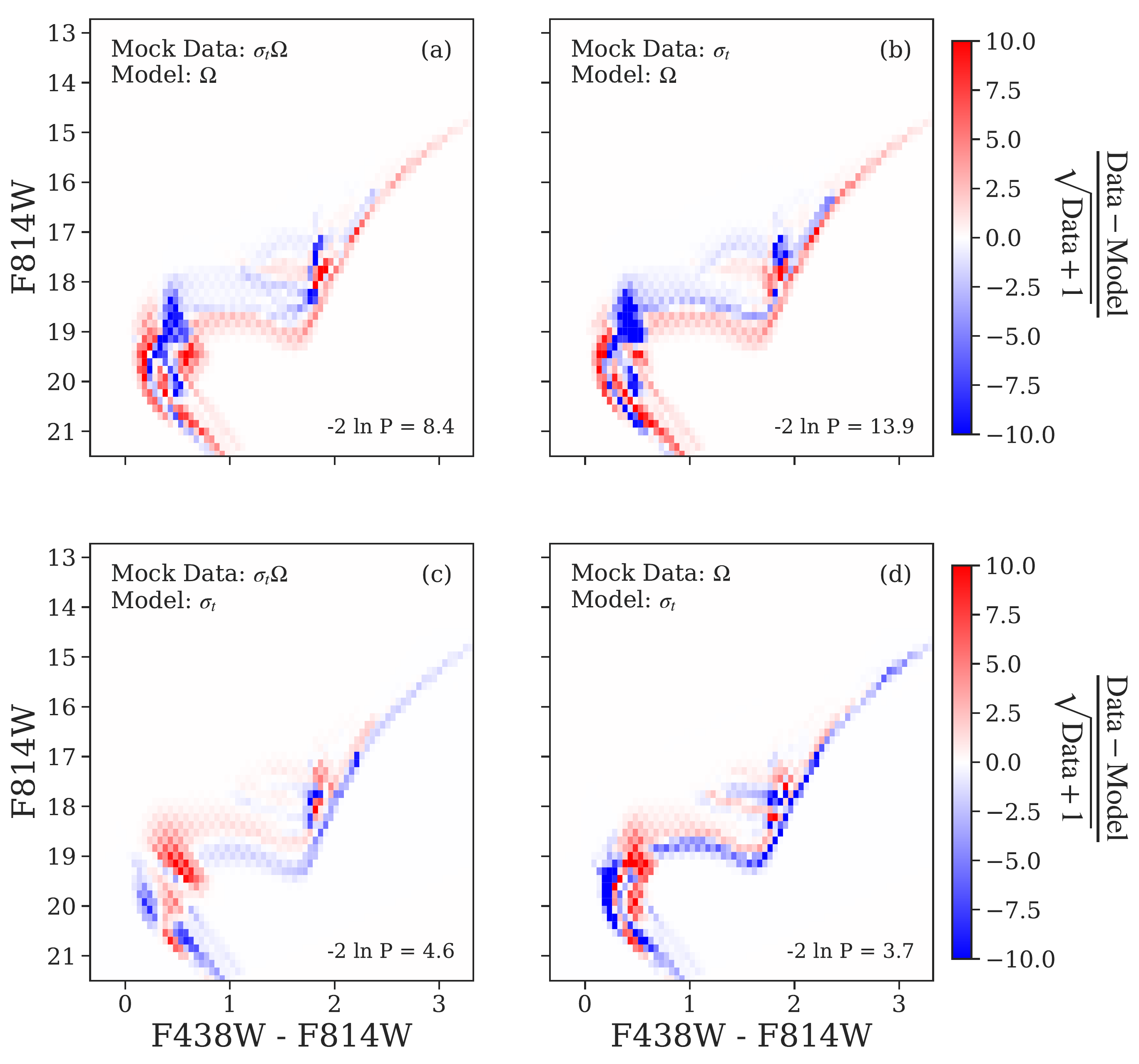}
    %\vspace{0.1cm}
    \caption{Selected residuals from the mock tests. The top row shows 
             results from cases where the $\Omega$-model was fit to either 
             of the other two models ($\sigma_t\Omega$ or $\sigma_t$). The 
             bottom row shows the same, but for the $\sigma_t$-model (fit 
             to either $\sigma_t\Omega$ or $\Omega$-mock data). Cases for the 
             $\sigma_t\Omega$-model, and all 1:1 fits (e.g., $\Omega$-model to 
             $\Omega$-mock data) were essentially zero throughout, and are 
             not displayed.}
\label{fig:mock_res}
\end{figure}

\subsection{Fitting to Data}
\label{ssec:fitmethod}

We build composite stellar populations 
according to three scenarios (below ``weights'' measure 
the total number of stars in a Hess diagram whose stars 
may be observations or simulations): 

\begin{enumerate}
    \item 
    \textbf{Gaussian log(Age) spread ($\sigma_t$-model)}: 
    This has a Gaussian SFH. More complex SFH are 
    imaginable, but would expand parameter space 
    further, and so are omitted for now. Here 
    $\vvc$ is restricted to 0.0 (non-rotating). 
    This model has three parameters 
    describing the overall weight, or number of stars 
    in the composite $\vvc=0.0$ population (i.e., 
    amplitude), plus the Gaussian log(Age) 
    distribution's mean and standard deviation.

    \item 
    \textbf{Non-parametric rotation rate distribution 
    ($\Omega$-model)}: This model considers 10 free 
    weights ranging from 0 to the total weight 
    of observed stars. Each free weight corresponds to 
    one of the 10 possible $\vvc$ populations. All 
    stars are assumed to have a single age, in this 
    model, which is also fit as an 11th free 
    parameter.
    
    \item 
    \textbf{Age spread with rotation 
    ($\sigma_t\Omega$-model)}: This model combines 
    the $\sigma_t$- and $\Omega$-models. A Gaussian 
    age distribution is allowed (mean age and 
    standard deviation are free parameters), as 
    are the 10 free weights for a non-parametric 
    rotation distribution, giving a total of 12 
    parameters. Like the $\sigma_t$-model, the age 
    distribution here is in terms of log(Age). 

\end{enumerate}

We measure the probability of models matching the 
data using Hess diagrams and a Poisson likelihood as 
described in \cite{AD2001}. Our fitting 
considers up to 10 independent weights 
for the density of stars at our 10 
values of $\vvc$, plus up to two more 
parameters describing the Gaussian age 
distribution's mean and standard 
deviation. We take this standard 
deviation of the Gaussian age distribution 
to represent the ``age spread''. So, given a 
derived mean log age $\mu_{\tau}$, we take 
the age spread to be 
$10^{(\mu_{\tau}+\sigma_{\tau})} - 
10^{(\mu_{\tau}-\sigma_{\tau})}$, with 
$\sigma_{\tau}$ being the standard deviation. 
We use Markov chain Monte Carlo (MCMC) to 
sample the probability distributions, 
determining the most likely parameter values for 
rotation rate weights, mean ages, and age 
spreads. We employ \texttt{emcee}'s affine-
invariant ensemble sampling algorithm 
(\citealt{DFM2013}). To initialize our 
ensemble of walkers, we chose randomized 
positions from a uniform distribution within 
$\pm 0.2$ dex of the chosen mean log age for 
a cluster, and within $\pm 0.05$ dex of the 
arbitrarily chosen initial age spread of 
$\Delta$ log age = 0.05, if applicable. In 
initializing the walker positions for the 
various $\vvc$ weights was done 
using a Dirichlet distribution. The 
reasoning behind this choice being that we 
desired these random initial positions to 
lie within the solution plane, such that all 
rotation rate weights add up to the total 
combined weight of all bins for the data. In 
other words, we set these positions 
in a way that preserves a cluster's total 
number of stars, rather than initializing in an invalid 
portion of parameter space where the total 
counts is far off from that of the data. Thus, 
our model for rotation distributions is a 
non-parametric model consisting of 10 free 
parameters; our eSF model is a Gaussian model 
described by its mean and standard deviation.

\begin{figure*}[!ht]
  \center
    \includegraphics[width=0.95\linewidth]{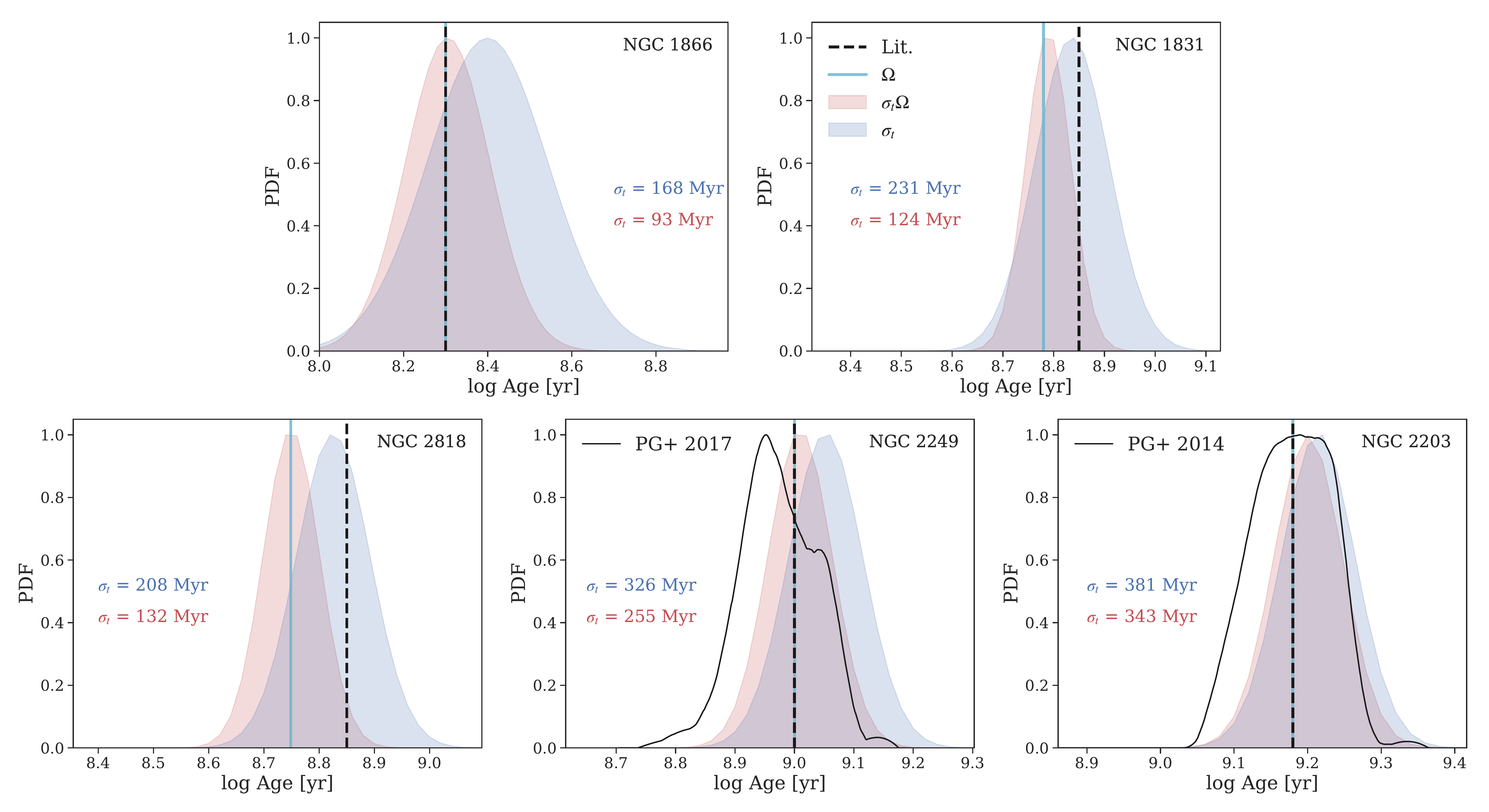}
    %\vspace{0.1cm}
    \caption{Derived ages according to the $\sigma_t\Omega$- and 
             $\sigma_t$-models (red and blue respectively). As 
             the $\sigma_t\Omega$-model allows non-zero rotation 
             rates, it typically finds a smaller age spread and 
             younger age than the $\sigma_t$-model, as discussed 
             in the text. Vertical black dashed lines indicate 
             ages cited from literature in Section \ref{sec:data}; 
             transparent cyan solid lines indicate the best-fit 
             age found by the $\Omega$-model for the respective 
             cluster. The black solid curves are pseudo-age distributions 
             from \cite{PG2014} and \cite{GGC2017} for NGC 2203 
             and 2249 (respectively) for comparison.}
\label{fig:clusterage}
\end{figure*}

\subsection{Mock Tests}
\label{ssec:mocks}

Here we present the results of mock tests 
examining accuracy in parameter recovery. 
These tests were carried out for our three 
scenarios of eMSTO presence under 
consideration: population age spread ($\sigma_t$-
model), stellar rotation distribution 
($\Omega$-model), or both 
($\sigma_t\Omega$-model). We generated mock 
data according to each of these scenarios 
using MATCH and applied our models to check 
that the input age and relative weights of 
populations at various $\vvc$ 
were recovered. The mock data is generated 
with a mean log age 9.0 in all cases, 
metallicity of [Fe/H] = -0.40, 
$A_{\rm{V}} = 0.07$, akin to NGC 2249, 
as determined in \cite{MC2014}. We generate 
mock data according to each scenario, and the 
log(age [yr]) spread is $0.05$, while the input 
rotation distribution is a Gaussian centered 
at $\vvc=0.4$ with a standard deviation of 
$0.2$ dex in applicable cases.

The recovered weights shown in Figure 
\ref{fig:mock_weights} find values near the 
inputs (black, hatched), and the truth is 
contained within errors in most cases, except 
the two shown in the bottom-left and -right 
panels (d) and (f). These two cases correspond 
to fitting the $\Omega$-model to mock data 
created with either the $\sigma_t$- in panel (d), 
or $\sigma_t\Omega$-model in panel (f). 
In both of these cases, the model forms 
roughly a bimodal distribution of rotation 
rates; both of these cases are the 
$\Omega$-model fit to mock data containing an 
age spread. Thus, we expect that our 
$\Omega$-model sees an age spread as a 
bimodal distribution of rotation rates. The 
reason for this is that fast rotators can 
enhance the eMSTO spread with more dramatic 
gravity darkening effects, as shown in e.g., 
\cite{BdM2009}, \cite{BH2015c}, and 
\cite{SG2018}. Additionally, our MIST-based 
rotating models mostly become redder as 
$\vvc$ rises, populating a reddened MSTO, 
allowing a greater TO spread but leaving 
behind a de-populated blue MSTO. Low rotation 
rates refill the bluer side of the MS when 
added to the full ensemble of stars. The 
bimodal nature of rotation seems to arise 
from fast rotators being used in an effort to 
fit the eMSTO spread, leaving a de-populated 
blue MSTO, alongside slow rotators 
counteracting this offset and populating the 
red MSTO. So, the properties of fast and 
slow rotators appear to drive our models to 
favor the presence of both in explaining an 
age spread, or otherwise broad MSTO. Unless 
the underlying rotation distribution in the 
data is something specific, like a Gaussian, 
we expect to see a bimodal distribution arise 
in order to model the width of the eMSTO. This 
is shown in panel (e) for $\Omega$-model fit 
to itself when the input rotation distribution 
is a Gaussian. The full model (i.e., the 
$\sigma_t\Omega$-model) is capable of recovering 
the input rotation rate distribution in all 
cases, as may be seen in panels (a), (b), (c). 
The additional degrees of freedom allowed by 
the Gaussian age spread in the $\sigma_t\Omega$
-model removes the necessity for a bimodal 
rotation rate distribution.

Figure \ref{fig:mock_res} shows residuals for 
our mock tests, again omitting 1:1 comparisons 
and only showing cases where the comparisons 
produced interesting residuals. In panel (b) 
of Figure \ref{fig:mock_res}, one may see that 
the $\Omega$-model does not match the smooth 
variation of stellar densities created by a 
Gaussian age spread (as in the $\sigma_t$-model). 
It does better matching the $\sigma_t\Omega$-mock, 
seen in panel (a) of Figure \ref{fig:mock_res}, 
(i.e. it acquires a higher likelihood), but 
similarly misses the continuous morphology of 
a Gaussian age spread; the placement of rotating 
models on the CMD imposes a combination of 
relatively specific morphologies, discretized 
according to a corresponding rotation rate. Thus, 
there is a low chance that each of these 
morphologies matches the more ambiguous morphology 
of a Gaussian age spread, creating distinct 
features in the residuals and relatively poor fits 
when the two try to match each other. 
Finally, the bottom row of Figure 
\ref{fig:mock_res} shows the $\sigma_t$-model 
fit to the $\sigma_t\Omega$-model in panel (c) and 
to the $\Omega$-model in panel (d). The $\sigma_t$
-model is generally capable of achieving higher 
likelihoods than the $\Omega$-model shown in the 
top row, but it still has difficulty reproducing 
the densities of stars at non-zero rotation rates, 
leading to the features shown in the residuals. 
This indicates that the $\sigma_t$-model possesses 
enough ambiguity to smooth out inconsistencies and 
produce a higher fit statistic with disregard to 
the presence of a distribution of rotation rates. In 
comparing panels (a) and (c), it is noticeable that 
the sub-giant branch (SGB) of the $\sigma_t\Omega$-
mock data is better matched by the $\Omega$-model.

\section{Results}
\label{sec:results}

We provide results in this section. Section 
\ref{ssec:resparam} provides a description of 
the derived rotation rate and 
ages distributions. Section \ref{ssec:res} 
presents the residuals between our best-fit 
models and data. Section \ref{ssec:dage} discusses 
our resulting $\Delta$(Age)-age trend with 
comparison to previous studies.

\begin{figure*}[!ht]
  \center
    \includegraphics[width=0.95\linewidth]{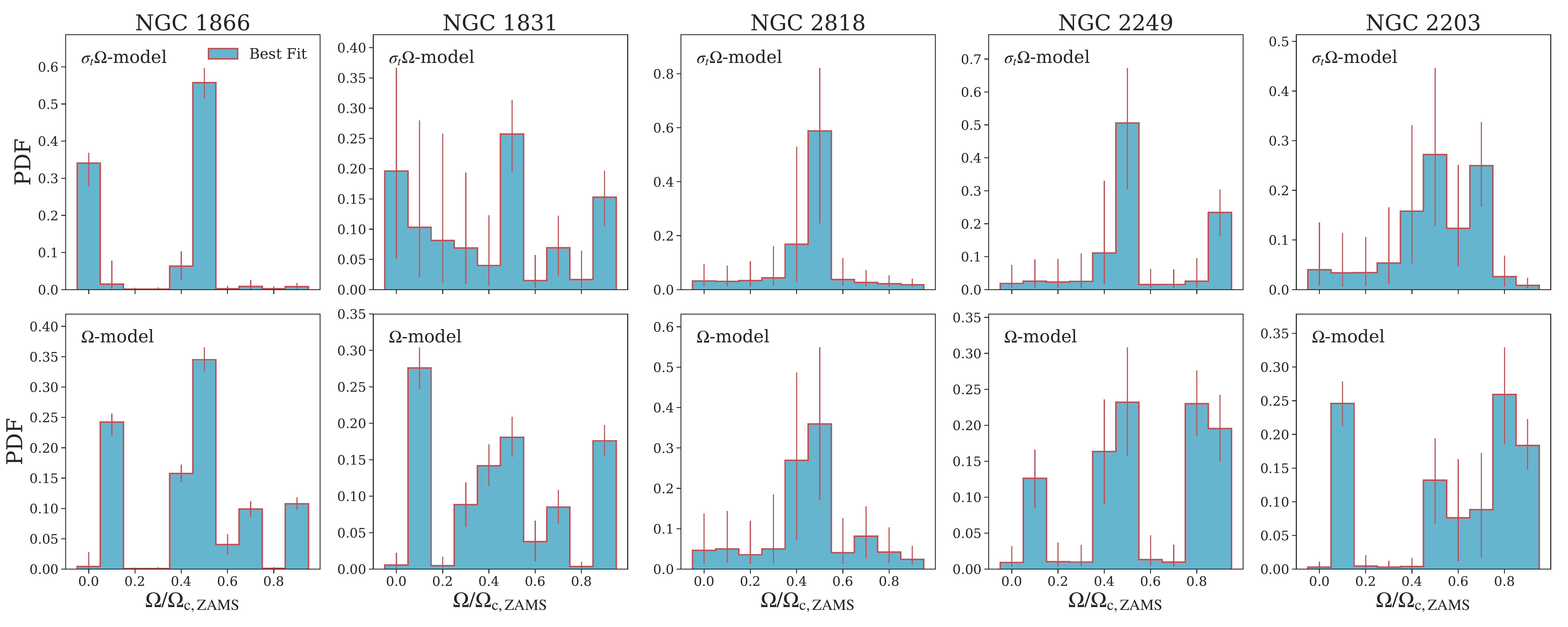}
    %\vspace{0.1cm}
    \caption{Rotation rate distributions resulting from our fits to (in columns 
             going left to right) NGC 1866, 1831, 2818, 2249, and 2203. The 
             top row shows results for the full $\sigma_t\Omega$-model, and 
             results for the $\Omega$-model are in the bottom row. The 
             $\sigma_t\Omega$-model tends to favor a single population of 
             moderate-fast rotators, while the $\Omega$-model tends to favor a 
             bimodal distribution of fast and slow rotators, with few 
             intermediate rates, similar to some recent observations of eMSTO 
             stars.}
\label{fig:rotdists}
\end{figure*}

% discussion of results/plots

\subsection{The Age and $\vvc$ Distributions}
\label{ssec:resparam}

Derived Gaussian age distributions for the 
$\sigma_t$- (blue shaded region) and 
$\sigma_t\Omega$-models are shown in Figure 
\ref{fig:clusterage}, with the best-fit age 
of the $\Omega$-model as a solid cyan line. 
Black dashed lines show the literature ages 
described in Section \ref{sec:data}. For NGC 
2203 and 2249, black solid curves indicate 
the ``pseudo-age'' distributions determined 
by \cite{PG2014} and \cite{GGC2017}, 
respectively. It is also mentioned in those 
works that the pseudo-age distributions are 
broader than what photometric errors allow, 
and so are not spurious in that manner.

Briefly, the pseudo-age distribution is 
one method of determining the age spread on 
the eMSTO. It is created with a 
parallelogram that encloses the width of 
the eMSTO. The colors and magnitudes of 
stars within this parallelogram are 
translated into an age distribution. This 
is done by taking the ages predicted by 
stellar models of these stars. Like our 
Gaussian age distributions shown in Figure 
\ref{fig:clusterage}, the pseudo-age 
is not reflective of the SF history of the 
cluster on its own (\citealt{PG2014}), but 
rather emulates the distribution of ages 
that may be present in the eMSTO at the 
time of observation.

In comparison to our derived age distributions, 
the pseudo-age distributions have multiple 
peaks, with the strongest amplitude at younger 
ages than we find. This may be due to the 
different models used (Padova in 
\citealt{PG2014} and SYCLIST in 
\cite{GGC2017}, while ours are MIST-based).
Additionally, it could be due to our age 
spread model only allowing a single peak, 
causing it to compromise for a peak in 
between the multiple peaks found by 
\cite{PG2014, GGC2017}. In spite of these 
offsets in mean age, the widths of the 
pseudo-age and Gaussian age distributions 
are comparable, suggesting that both methods 
predict similar age spreads.

However, a common trend seen with all 
clusters is that the $\sigma_t\Omega$-model 
predicts a smaller age spread. This is 
expected, as stellar rotation also 
contributes to the eMSTO morphology with 
this model. Gravity darkening can introduce 
substantial color variations and eMSTO width, 
as demonstrated in Figure \ref{fig:fakephot}. 
Furthermore, it may be seen that the inclusion 
of stellar rotation tends to reduce the 
predicted mean age.

Due to the reddening effect of gravity darkening, 
it is also expected that stellar rotation would 
reduce the mean age. Age is primarily determined 
by the CMD position of the MSTO, which is fixed 
by the data. As stellar rotation tends to redden 
stellar models, selecting a younger age 
counteracts this effect by making the rotating 
stars blue again. Hence, stellar rotation is seen 
to derive a younger age than non-rotating models 
(the $\sigma_t$-model) in Figure \ref{fig:clusterage}. 
The ages found by the $\Omega$- and $\sigma_t\Omega$-
models are either similar or coincide.

Although the $\sigma_t\Omega$- and $\Omega$-model 
agree on cluster ages, the derived $\vvc$ 
distributions shown in Figure \ref{fig:rotdists} 
highlight where these scenarios disagree. The 
top row shows the distributions found by the 
$\sigma_t\Omega$-model, while results for the 
$\Omega$-model are on the bottom row. In all 
cases, the $\Omega$-model finds a more 
distinct population of fast (e.g., $\vvc 
\geq 0.5$) and slow rotators ($\vvc < 0.5$). 
In contrast, the $\sigma_t\Omega$-model 
generally finds a smaller presence, or lack 
of slow rotators. 

\begin{figure*}[!t]
  \center
    \includegraphics[height=0.9\textheight]{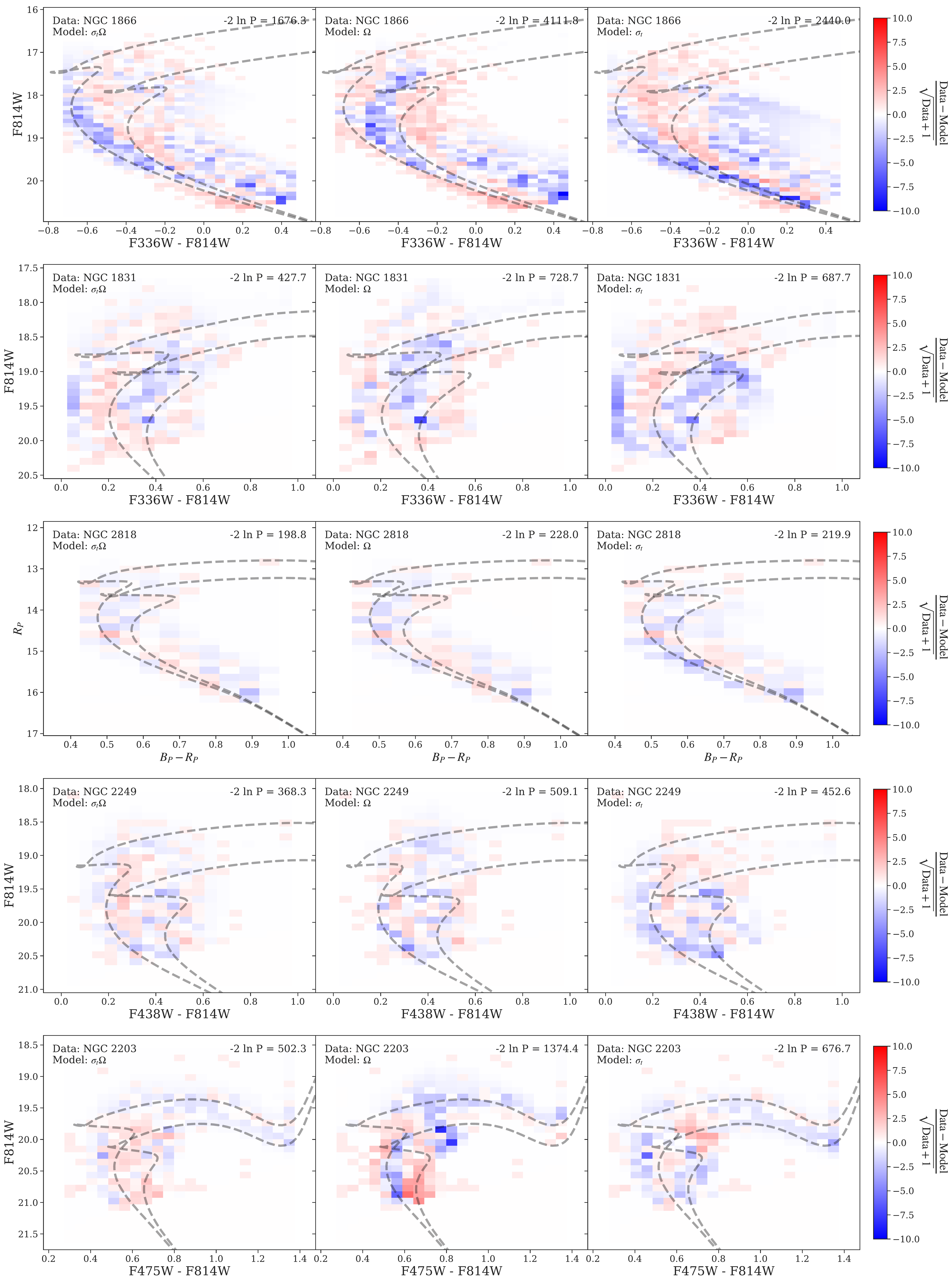}
    %\vspace{-2pt}
    \caption{Residuals for the fits to real data. Each row 
             pertains to a star cluster (top to bottom): 
             NGC 1866, 1831, 2818, 2249, 2203. Each column 
             contains results for a particular model (left to 
             right): $\sigma_t\Omega$-, $\Omega$-, $\sigma_t$-model. 
             Blue pixels correspond to where the model 
             overestimates the data, and red pixels to where it 
             underestimates. Fit statistics are annotated in the 
             upper right corners of each panel. The black dashed 
             lines are non-rotating isochrones at ages chosen to 
             roughly span the observed eMSTO width of each 
             cluster, for reference.}
\label{fig:residuals}
\end{figure*}

The $\Omega$-model's rotation rate 
distributions agree with observations more 
than the $\sigma_t\Omega$-model in 
this context. They qualitatively reproduce 
the observed fast and slow rotators found by 
e.g., \cite{AKD2017, BN2018}. In the 
observations, slow rotators reside blueward, 
while fast rotators lie redward. The stellar 
models capture this behavior as well, as 
seen in Figure \ref{fig:fakephot}; gravity 
darkening tends to redden fast rotators, 
causing the two populations to occupy 
distinct color spaces in the CMD. The two 
populations are necessary in the stellar 
rotation scenario so that the full 
blue to redward extent of the eMSTO is 
reproduced.

The two populations are not required in 
the $\sigma_t\Omega$-model. The Gaussian 
age spread can compensate for a lack 
of slow rotators. This does lead to a 
clear lack of slow rotators in 
NGC 2203 and 2249 (the oldest clusters). 
For these clusters, the predicted lack of 
slow rotators is not in line with 
recent findings for younger clusters.

\subsection{Residuals}
\label{ssec:res}

Figure \ref{fig:residuals} shows the residuals 
of the data compared to the best-fit 
models for the $\sigma_t\Omega$- (left 
column), $\Omega$-

\noindent
\clearpage
\noindent
(middle), and 
$\sigma_t$-model (right). Dashed black lines 
show non-rotating MIST isochrones placed 
to trace the eMSTO width (similar to the 
isochrones in Figure \ref{fig:fakephot}). 
Each row corresponds to a cluster, and the 
global likelihoods (-2lnP) are kept in 
the upper right corner of each panel. While 
the $\sigma_t\Omega$-model may not reproduce 
the rotation rate distributions one might 
expect, it is the best-fit model overall, 
considering the global likelihoods. 

In NGC 2249, 2818, and 1831, the 
likelihoods of all three scenarios are 
comparable, and the best matched clusters 
on the basis of the residuals. Meanwhile, 
NGC 2203 and 1866 (the oldest and youngest 
clusters) are the worst matched. In the 
latter two cases, performance between the 
three scenarios is more disparate. The 
$\sigma_t\Omega$-model formally achieves he 
best-fit with all clusters (perhaps 
unsurprisingly given more free parameters), 
but clearly so with these two 
clusters in particular. None of the 
residuals are clean, showing that all 
scenarios produce imperfect 
fits, although in different ways.

%%\clearpage
%%\noindent
Inclusion of an age spread does tend to 
deliver a higher likelihood. At the same 
time, the residuals show that the $\sigma_t$- 
and $\sigma_t\Omega$-models often create eMSTOs 
that are broader than the data. This is seen 
both by pixels over fit by these models tend 
to lie outside of the bounding isochrones 
that roughly trace the eMSTO. The $\Omega$
-model tends to over predict within the 
isochrones, without the extended behavior of 
the Gaussian age spread models.

The $\Omega$-model can find comparable 
likelihoods to the age spread models, but 
also finds mismatches. In fact, it is formally 
the worst fit in all cases. Yet, it does appear 
to match the eMSTO extent well qualitatively, 
suggesting that it fails moreso in getting the 
precise stellar densities correct. In other words, 
the $\Omega$-model appears to reproduce the data
morphologically, though it does not show a strong 
statistical advantage over the other two models.

\begin{figure}[!t]
  \center
    \includegraphics[width=0.95\linewidth]{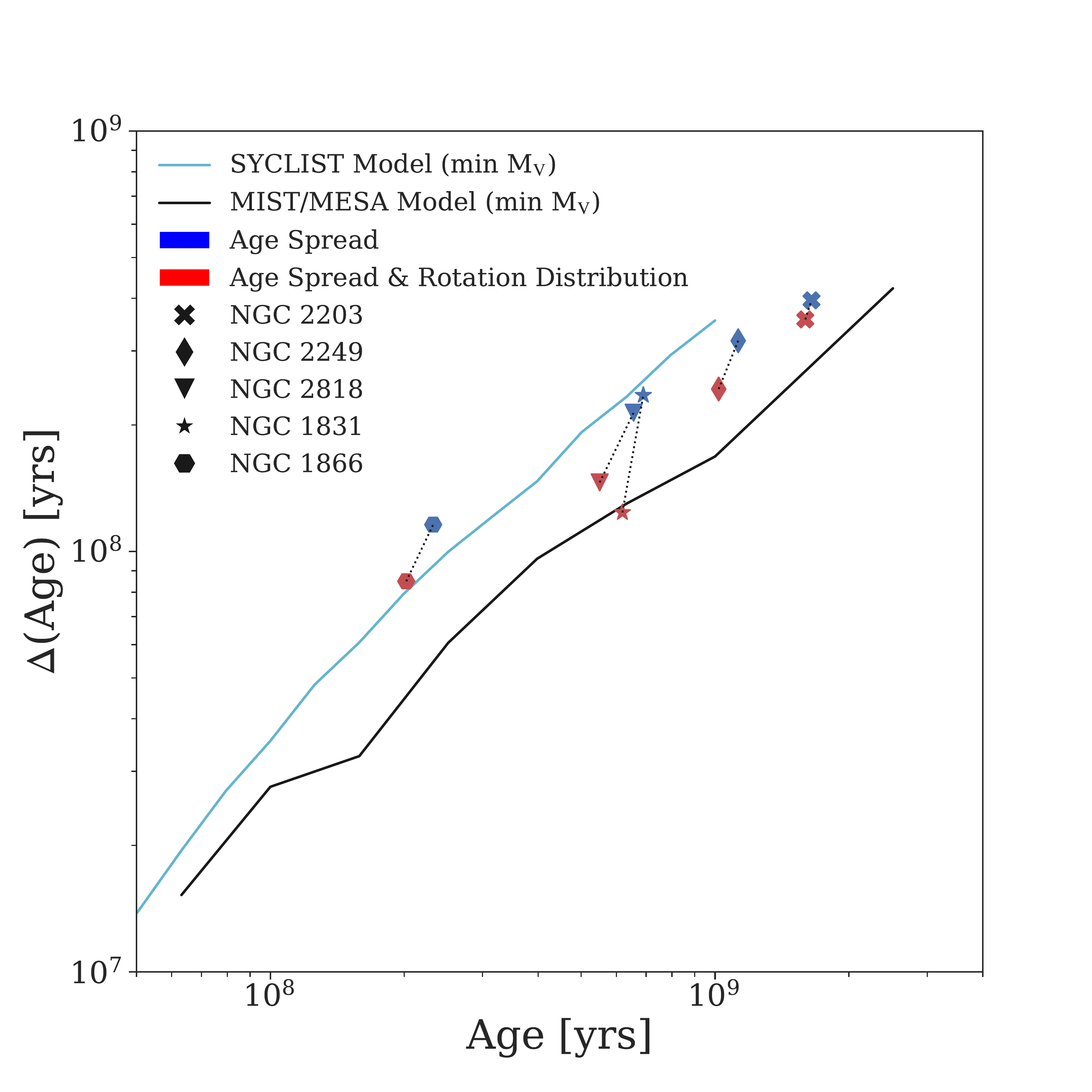}
    %\vspace{0.1cm}
    \caption{Blue and red points show derived age and age spreads 
             according to the $\sigma_t$-model and $\sigma_t\Omega$
             -models, respectively. Also shown lines comparing 
             the same trend found via the SYCLIST models 
             in \cite{NF2015} (magenta solid), with our MIST-
             based model prediction (black solid).}
\label{fig:nh15}
\end{figure}

\subsection{The $\Delta$(Age)-Age Trend}
\label{ssec:dage}

The $\Delta$(Age)-age trend, such as 
the one derived here and shown in 
Figure \ref{fig:nh15}, has been a 
central point in eMSTO studies such as 
\cite{NF2015}. The points in Figure 
\ref{fig:nh15} mark age spreads and 
ages found by the $\sigma_t$- (blue) and 
$\sigma_t\Omega$-models (the latter in 
red). For a series of ages, 
\cite{NF2015} determined the effective 
$\Delta$(Age) that a non-rotating 
GENEC-based, SYCLIST isochrone would need 
in order to match a $\vvc=0.5$ isochrone. 
They selected distinct points on each 
isochrone, and determined how much the 
non-rotating isochrone's age needed to 
shift from its original value in order 
to match the CMD position of the 
rotating isohrone. Remarkably, they found a 
$\Delta$(Age)-age trend that agreed 
with measured eMSTO widths from a 
range of studies. The lines in Figure 
\ref{fig:nh15} show such model predicted 
trends. These $\Delta$(Age)-age trends 
imply that the effect of stellar 
rotation can mimic an age spread.

The magenta line in Figure \ref{fig:nh15} 
is the trend formed by their second 
isochrone point, ``$M_V$ at MSTO''. We 
performed the same analysis to see what the 
MIST models predict, and this trend is 
shown as the solid black line. The MIST 
models produce a shallower trend. In 
\cite{SG2018}, we found that MIST models 
predict a smaller apparent age spread as a 
result of rotation when compared to SYCLIST, 
due to our weaker rotational mixing. This 
causes the luminosities of our rotating and 
non-rotating isochrones to separate less 
than SYCLIST models, leading to a smaller 
effective ``age spread''. We also found 
that gravity darkening appeared to be the 
strongest effect of stellar rotation in the 
MIST models.

The black and magenta lines in Figure 
\ref{fig:nh15} do not show the inclination 
angle dependence of rotationally induced 
eMSTO width. These lines were measured with 
the luminosity and temperature of the stellar 
models averaged over all inclination angles. 
Thus, this trend does vary slightly depending 
on the chosen inclination angle, but it remains 
that on average the MIST models predict a 
shallower trend than SYCLIST when measuring the 
eMSTO spread with this isochrone-based 
method. In a synthetic population (as opposed to 
single isochrones), the combined effects of a 
distribution of rotation rates and inclination 
angles allow the rotating MIST models to achieve 
comparable eMSTO widths to the observed 
$\Delta$(Age)-age trends. 

\begin{figure*}[!t]
  \center
    \includegraphics[width=0.95\linewidth]{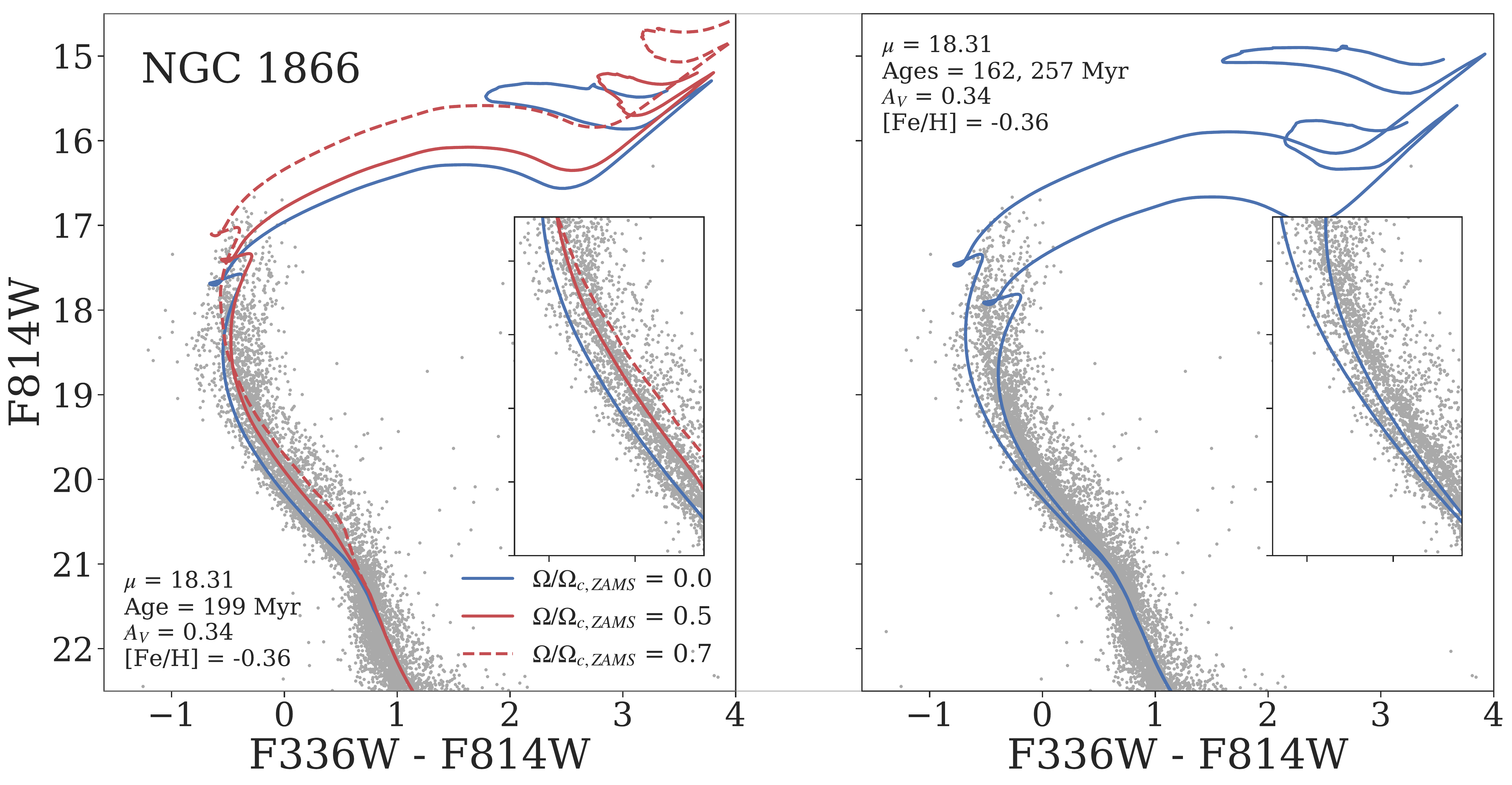}
    %\vspace{-2pt}
    \caption{Schematic demonstrations of the age spread/rotation 
    rate distribution contention in NGC 1866. \textbf{Left:} Non-rotating 
    and rotating ($\vvc=0.5, 0.7$) MIST isochrones show a reproduction of the 
    split MS (highlighted in the inset), although they do not reproduce the 
    CMD location of the bluest and reddest TO stars. The age here is the age 
    found by the best-fit $\Omega$-model. Here, $\vvc=0.5$ is the largest population 
    of fast rotators found by the best fit model, while $\vvc=0.7$ is included to 
    show how it aids in reproducing the eMSTO. \textbf{Right:} Non-rotating 
    isochrones varying in age show that eSF could in theory match the eMSTO 
    width, covering the bluest and reddest regions of the cluster TO. The 
    age range is chosen so that the isochrones roughly span the eMSTO width.}
\label{fig:n1866isos}
\end{figure*}

\begin{figure}[!t]
  \center
    \includegraphics[width=0.95\linewidth]{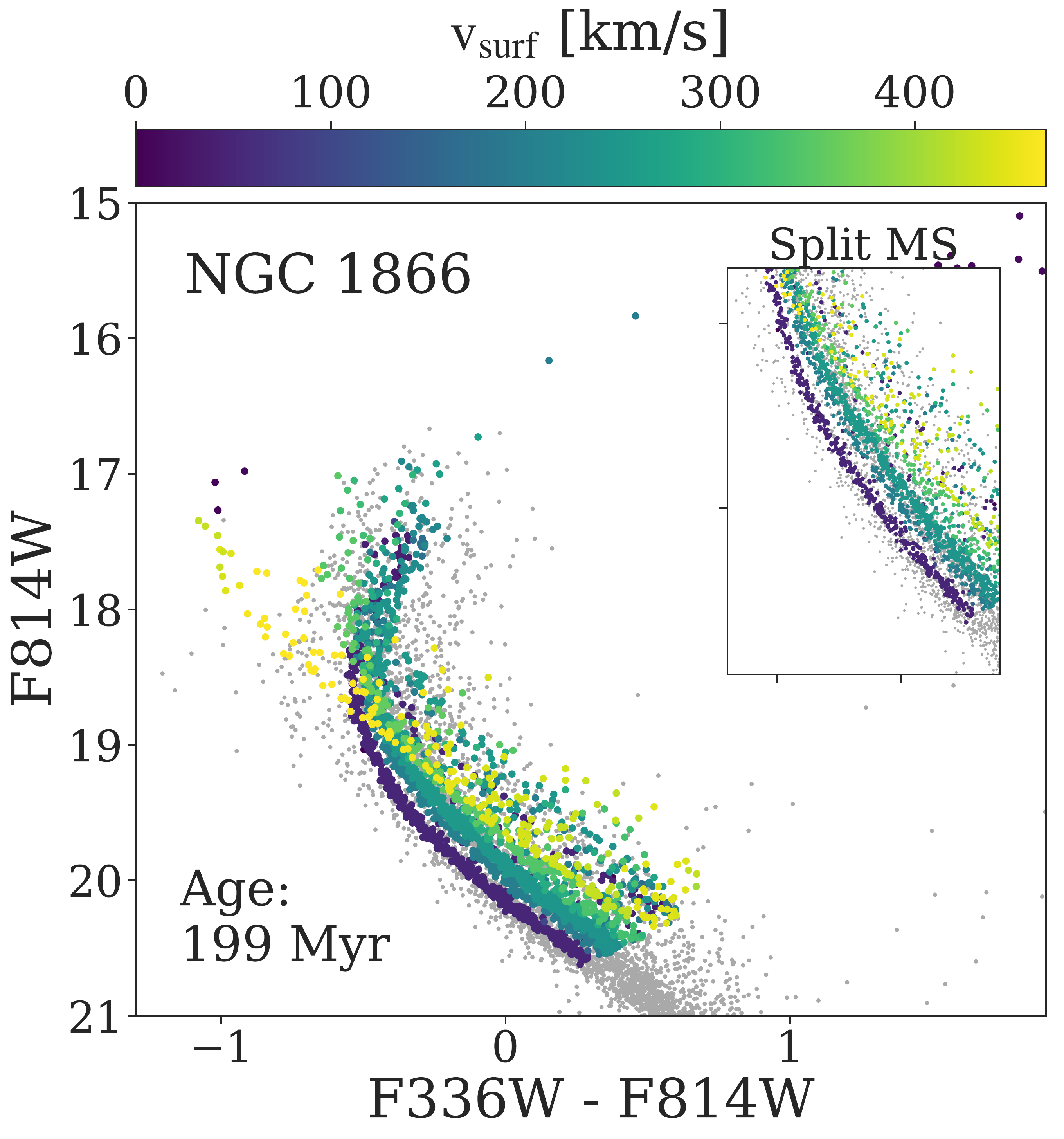}
    %\vspace{-2pt}
    \caption{Colored points are the 
    best-fit $\Omega$-model for NGC 1866, with its derived 
    age in the lower left. The inset focuses on the 
    split MS, showing that this model qualitatively 
    replicates the split MS with a distribution of 
    rotation rates, mainly slow rotators combined with 
    fast rotators at $\vvc=0.5$. Grey points are the data. 
    The yellow points are stars rotating near critical, 
    which in our modeling does not match the data; this 
    population appears to be spurious. The reddest TO 
    stars are also missed; these may be fast rotators 
    reddened by decretion disks (Be stars).}
\label{fig:n1866best}
\end{figure}

In the derived ages and age spreads of the 
$\sigma_t$- and $\sigma_t\Omega$-models, we see 
that the inclusion of stellar rotation does 
not reduce the age spread to zero. The age 
spreads derived by the $\sigma_t\Omega$-model 
are smaller than those found by the $\sigma_t$
-model. This is expected, as stellar rotation 
and age spread compete to explain the eMSTO 
morphology in the $\sigma_t\Omega$-model. The 
fact that the $\sigma_t\Omega$-model's age 
spreads still correlate with cluster 
age suggests that they trace a stellar 
evolution effect, rather than SF. The 
fact that the $\sigma_t\Omega$-model finds a 
reduced, but non-zero age spread may be 
indicative of missing physics in the rotating 
models, compensated for by an age spread.

\subsection{The Split MS of NGC 1866}
\label{ssec:splitms}
NGC 1866 has a split MS, in 
addition to its eMSTO, that warrants further 
analysis. This additional complexity may contribute 
to the relatively poor fit achieved for NGC 1866 (Figure 
\ref{fig:residuals}). The split MS has been argued 
to imply that star clusters host both eSF and a 
distribution of rotation rates. The fact that the 
$\sigma\Omega$-model achieves the best-fit here, 
and the $\Omega$-model the worst, serves to 
demonstrate how eSF may be compensating 
for an incomplete modeling of stellar evolutionary 
effects (e.g., binary mergers and decretion disks). 
In spite of the lower statistical likelihood of the 
$\Omega$-model (Figure \ref{fig:residuals}), we 
find that a coeval distribution of rotation rates 
is capable of reproducing the split MS and the 
eMSTO simultaneously, albeit with caveats.

Schematically highlighted in Figure \ref{fig:n1866isos} is 
the debate between whether the split MS 
is due to a rotation rate distribution, 
or whether eSF may be present. Figure \ref{fig:n1866isos}
displays content similar to Figure 10 
from \cite{APM2017}. A range of ages clearly 
does not reproduce the split MS (right panel). 
As also noted by \cite{APM2017, APM2018}, 
modeling the split MS appears to require 
fast and slow rotators in the cluster. This is 
shown for the simple case of a non-rotating and 
$\vvc=0.5$ isochrone ($\vvc=0.7$ is included as well to 
show how they aid in extending the MSTO). The split MS 
may be driven by bimodal rotation rate distributions, 
but $\vsini$ confirmations still need to be obtained. 
As plotted here, and as seen in Figure 
\ref{fig:rotdists}, the fast rotators are mainly 
stars rotating at $\vvc=0.5$, in contrast 
to the predictions of Geneva shown in 
\cite{APM2017, APM2018} where the red MS appears to 
be comprised of stars rotating near critical velocity.

The right panel of Figure \ref{fig:n1866isos} shows that a range of 
stellar ages can aid in spanning the bluemost and 
redmost regions of the eMSTO. This was also shown in 
the corresponding Figures 10 and 11 of \cite{APM2017}, 
where isochrones at several rotation rates and multiple 
ages are plotted together. This is one example of how 
eSF in models (e.g., $\sigma\Omega$) may optimize a 
fit to the data, but it is uncertain whether the invoked 
age spread is physical. As mentioned, there is sparse 
empirical evidence for eSF, and alternatives are known 
(although, see recent work by \citealt{GCosta2019a}, 
finding possible evidence for eSF in NGC 1866 via 
modeling of its Cepheid stars). 

Our best fit $\Omega$-model (which obtains the worst 
likelihood) for NGC 1866 is shown in Figure 
\ref{fig:n1866best}, with an inset focusing on the 
split MS. The recovered rotation rate distribution 
for NGC 1866 is indeed bimodal (for both the 
$\sigma\Omega$- and $\Omega$-model), and the model 
qualitatively matches the split MS. The eMSTO is 
matched in some respects, e.g., that it predicts a 
mixed population of slow and fast rotators, as 
observed spectroscopically by \cite{AKD2017}. In 
other areas, a rotation rate distribution is 
inconsistent with the data as modeled 
here. Particularly, it misses stars blueward (i.e., 
about F336W - F814W $<$ -0.65 mag) and redward (about 
F336W - F814W $>$ -0.35) in the eMSTO, but these areas 
of the CMD are known to be affected by complex 
stellar evolution effects, e.g., interacting binaries and 
Be stars, respectively.

For instance, our models do not include properties such 
as decretion disks, nor binary merger products. 
The population of fast rotators extending blueward may 
not reproduce the findings of \cite{AKD2017}, 
and could be a spurious result. Increasing the binary 
fraction to 50\% and removing the possible ``blue 
stragglers'' does not influence the presence of 
these near critical rotators in the best fit; so, their 
presence appears to aid in matching stellar densities 
along the MS, rather than the eMSTO, with these near 
critical rotators extending blueward of the TO as a 
possible side effect. On the matter of reddened, fast 
rotators, \cite{BN2017} found 
evidence of a high number of H$\alpha$ emitters 
(suspected Be stars) throughout the eMSTOs of these 
young clusters. A number of which lie in this redward 
region. Though, \cite{MC2017} examined whether H$\alpha$ 
emitters account for the redward extension of the 
MSTO in the young cluster NGC 1850; they found that 
H$\alpha$ emitters could not in NGC 1850, in comparing 
their data with SYCLIST stellar models. See the modeling done by 
\cite{GA2018} for more on Be stars and further examples 
in young clusters. As for the blueward extent, \cite{CL2019} 
show some examples of the blue straggler populations that 
might exist in Magellanic Cloud clusters, but \cite{DE2019} 
warn that such stars may also be unaccounted for field 
contaminants. Presently for our NGC 1866 data, these stars 
do not reside in the field, and so appear to be cluster 
members.

Rather than eSF, it is possible that the 
redmost and bluemost populations of eMSTO stars (where 
the $\Omega$-model does not reproduce the data well, thus 
calling to question if this is evidence of eSF) are 
affected by decretion disks and binary interaction. 
These features are also missed by the $\Omega$-model 
in NGC 1831, 2818, and 2249. The 
$\sigma\Omega$-model may compensate for these missing 
factors by invoking an age spread. We will need to 
improve our stellar modeling, or obtain further 
data on what sort of stars these are before we can say 
for sure.

\section{Discussion \& Conclusions}
\label{sec:conclusions}

In this paper, we have compared the 
statistical ability of three proposed 
scenarios to explain the eMSTO 
morphology: 1) eSF, 2) a rotation rate 
distribution, and 3) both combined. In 
analyzing the statistics, we also 
considered whether the three scenarios 
could reproduce observed properties of 
eMSTOs, such as their rotation rate 
distributions. eSF remains a possibility 
based on this analysis, 
but it seems unlikely in our interpretation. 
The results highlight that a 
distribution of rotation rates is 
capable of solely accounting for 
eMSTO morphology and observed 
populations of fast and slow 
rotators in the eMSTO. Additionally, a 
distribution of rotation rates may 
simultaneously account for the split 
MS, and the eMSTO of the roughly 200 
Myr cluster NGC 1866, given current 
uncertainties in stellar modeling. Yet, the 
$\sigma_t$- and $\sigma_t\Omega$-
models formally achieve the highest 
likelihoods. Whether or not 
these age spreads are physical is put 
into question by our residuals. All 
three scenarios over/under-predict in 
different areas. The youngest and 
oldest clusters (NGC 1866 and 2203) 
are the worst fit clusters, but these 
are also the only cases where an age 
spread appears to significantly 
outperform the $\Omega$-model. 
Incidentally, these two cases are 
both affected by quite uncertain and 
unaccounted for aspects of stellar 
rotation: magnetic braking in 
NGC 2203, and Be stars in NGC 1866. In 
each case, the $\sigma_t\Omega$- 
and $\sigma_t$-model show strong 
over predictions blue- and redward of 
the observed eMSTO. This suggests that 
the Gaussian age spread may optimize 
the fit in these cases, but create 
broader eMSTOs than the bulk of the data 
to do so. In contrast, the $\Omega$
-model tends to find mismatches within 
the observed eMSTO region, while not 
producing such broad artifacts. It is known 
that our models are incomplete (e.g., 
in not modeling certain effects like the 
decretion disks of very fast rotators and 
in approximating rotational and 
convective mixing with a 1D framework). 
Thus, it seems plausible that a Gaussian 
age spread is compensating for missing 
aspects. In this section we discuss 
caveats, uncertainties, and suggestions 
for future work.

\begin{figure*}[!ht]
  \center
    \includegraphics[width=0.95\linewidth]{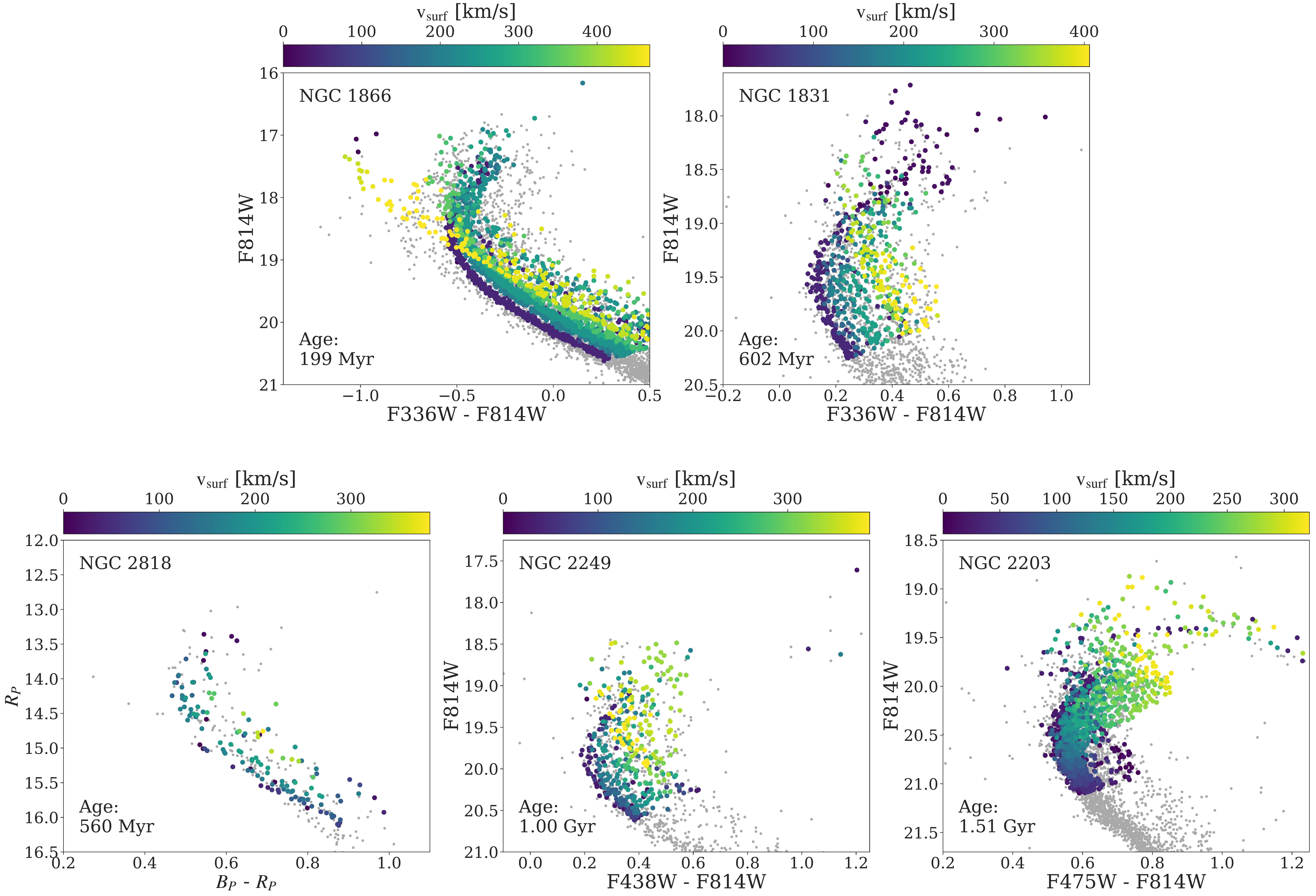}
    %\vspace{0.1cm}
    \caption{Each panel shows the best-fit $\Omega$-model (colored 
             points) for each cluster, overlaid on top of the data 
             (gray crosses). Similar to observations, our models 
             find a blueward population of slow rotators, while 
             fast rotators tend to lie redward. The color scale 
             corresponds to the average surface velocity of our 
             models. The best-fit age found by the $\Omega$-model 
             is shown in the panels.}
\label{fig:rotblend}
\end{figure*}

Figure \ref{fig:rotblend} shows 
the best-fit $\Omega$-models for each 
cluster. Grey crosses show the data, 
colored points show the best-fit 
$\Omega$-models, with colors mapped to 
the average surface velocity of the 
stellar models. Visually, the 
$\Omega$-model provides a good match, 
but is not perfect. For instance, models 
miss the redward MSTO of NGC 1866 which 
could be due to not including decretion 
disks that can further redden rotating 
stars (see Section \ref{ssec:splitms}). 
It is worth highlighting that NGC 2818 
contains a relatively low number of stars, 
so the model shown in Figure 
\ref{fig:rotblend} is subject to 
stochasticity when stars are drawn from the 
IMF. On multiple draws, the eMSTO of NGC 2818 
is visually reproduced more or less well as 
a result; it can become narrower or broader 
on subsequent draws. We have chosen a draw 
that matches the data fairly well. In NGC 
2203 and NGC 2249, the redward shelf-like 
structure roughly at magnitudes of 20 and 
19.3 for each cluster, respectively, is only 
matched by fast rotators (Figures 
\ref{fig:fakephot} and 
\ref{fig:rotblend} show this), but is 
over predicted in each case (see Figure 
\ref{fig:residuals}) with the 
$\Omega$-model. It is conceivable that 
missing physics related to stellar 
rotation, such as magnetic braking, 
could lead to model mismatches here.

Additionally, mismatches could come 
from associated effects, like stellar 
binary evolution. However, the effect 
of binary interaction may be minimal 
(\citealt{WY2011, ZL2016}) in 
intermediate age clusters, in 
comparison to the effects of 
rotation or a possible age spread. 
Though we have included unresolved 
binaries in our models, we have not 
incorporated the effects of binary 
interactions, which could affect 
rotation rates via tidal forces, or 
directly impact evolution through 
mass transfer, possibly producing objects like 
blue stragglers. Observations of 
eMSTO stars focused 
on determining their binary status 
could help shed light on the 
importance of these effects.

The origin and termination of 
rotation rate distributions is 
currently undetermined. Braking of 
stellar rotation via various mechanisms 
may be a source of influence. At the older 
end of eMSTO hosting star clusters, 
magnetic braking affects stars with 
convective envelopes. Such stars have 
masses $<1.8\msol$ or so, and start to 
arrive on the MSTO near 1 Gyr. 
Disappearance of the eMSTO appears to 
occur for clusters older than about 2 Gyr, 
where stars now possess larger convective 
envelopes and much stronger magnetic 
braking, consequently driving all TO stars 
in these older clusters towards slow 
rotation rates. E.g., \cite{SM2018a} found 
NGC 1978 to host an MSTO that is consistent 
with no spread within the 
observational errors. Although, the 
age of eMSTO disappearance may be 
metallicity dependent, as early work 
by \cite{CG2019} appears to suggest. Going 
towards younger (than about 600 Myr) 
clusters, such as NGC 1866, the 
split MS becomes a prominent feature in 
addition to the eMSTO. The origin of the 
split MS could be linked to a braking 
mechanism (as \citealt{FDAn2017, FDAn2018} suggest 
with tidal braking) that makes fast rotators 
transition to slow rotators, producing the 
observed bimodal MS in these young clusters. Just 
as well, at this point it is unclear if this bimodal 
distribution is an imprint of the initial velocities 
that stars may be born with.
Neither magnetic braking, nor binary evolution 
are modeled here, but both appear to be integral 
in understanding how rotation rate distributions 
evolve with time in these clusters.

Furthermore, several inconsistencies 
between the age spread scenario and CMD 
structure are relevant to whether 
significant age spreads are physical 
phenomena (\citealt{ICZ2018} gives an 
overview). Figure \ref{fig:fakephot} 
shows a broad SGB is expected in intermediate 
age clusters if eSF is present, while a 
distribution of rotation rates predicts a 
narrow SGB (also see \citealt{BdM2009}). 
However, models showing a narrow SGB are those 
of \cite{BdM2009}, which lacked interior 
rotational fuel mixing, and ours, which 
possess relatively weak rotational mixing. 
This is in comparison to either an age 
spread, or models with stronger rotational 
mixing, as those used in \cite{NF2015}, 
which may produce a broad SGB at these ages. 
A broad SGB does not show up clearly in 
the observations of NGC 2203 or 2249 used 
here. \cite{LdeGD2014} found age spreads 
were inconsistent with the SGB of NGC 1631, 
while \cite{PG2015} found that the SGB 
structure could be consistent with an age 
spread in some cases. The absence or 
presence of a broad SGB could be a useful 
determinant in constraining the physics at 
hand. 

Another important diagnostic could be the RC. 
The RC (excluded from our fits) should be 
broadened in intermediate age clusters, 
generally according to some effect (e.g., 
rotation rate distributions or an age spread) 
that can create a spread in stellar mass within 
a cluster. \cite{LG1999} laid out the physics 
of the extended RC; it requires that a 
range of stellar masses exists in the cluster, 
such that some RC stars developed degenerate 
He cores, while others were massive enough to 
bypass this. A range of ages is capable of 
creating this spread in stellar mass (see e.g., 
\citealt{PG2014, PG2015}), although rotating 
models have not been widely tested as an 
alternative. We aim to study the eRC according 
to the stellar rotation scenario in upcoming work. 
Alongside the eMSTO, studying the SGB and RC 
structure will inform a more complete picture 
of the role of stellar rotation, and could reveal 
missing physical ingredients within the models.

The physical basis 
of eSF is questioned further by the 
$\Delta$(Age)-age trend (Figure 
\ref{fig:nh15}). \cite{GGC2017} and 
\cite{APM2018} found that age spreads 
combined with a distribution of 
rotation rates provides a better 
reproduction of data than the latter 
acting solely. However, as 
\cite{BN2018} alluded to in their 
Figure 4, age spreads still appear to 
correlate with cluster age. This 
suggests that the age spreads 
determined in those cases, and here 
with the $\sigma_t\Omega$-model emanate 
from a stellar evolution effect, rather 
than true eSF. In general, the ability 
of stellar rotation to largely account 
for eMSTO structure on its own (as 
shown in Figure \ref{fig:rotblend}) 
may suggest that residual 
mismatches are signs of imperfect 
stellar modeling, rather than a true 
age spread. 

In our interpretation, 
a distribution of rotation 
rates appears to be the overall most 
physically motivated explanation for the 
eMSTO phenomenon. We 
can not rule out eSF here, as it does 
aid in creating a better fit to the data. 
However, the age spread may also be 
compensating for known 
missing ingredients in the rotating models, 
such as braking (tidal or magnetic), Be 
star disks, effects of binary interaction, 
and uncertainties in 1D convective mixing. 
This contention may resolve as stellar models 
improve. Direct comparison of observed eMSTO 
$\vsini$ with stellar models (such as those 
in Figure \ref{fig:rotblend}) should provide 
strong constraints on the physics of 
stellar rotation. There is known 
uncertainty between the formalisms that 
are used to simulate stellar 
rotation. A detailed assessment of 
MIST- with GENEC-based models should 
illuminate the consequent range of model 
behaviors. The SGB and RC, in 
combination with the eMSTO, may provide 
an opportunity to study, constrain, and 
reconcile many complex, and as of now 
uncertain stellar processes within young 
clusters.

\section*{Acknowledgments}

SG acknowledges the National Science 
Foundation Graduate Research Fellowship 
under grant No. DGE1745303. CC 
acknowledges support from the Packard 
Foundation. This paper is based upon 
work supported by the National 
Aeronautics and Space Administration 
(NASA) under Contract No. NNG16PJ26C 
issued through the WFIRST 
Science Investigation Teams Program. 
Some of this material is based upon work 
supported by the National Science 
Foundation under Award No. 1501205. ICZ 
acknowledges support from NASA through 
Hubble Fellowship grant 
HST-HF2-51387.001-A, awarded by the Space 
Telescope Science Institute (STScI), which is 
operated by the Association of Universities 
for Research in Astronomy, Incorporated, for 
NASA, under contract NAS5-26555. BFW acknowledges 
support from NASA through grant AR-13901 
from STScI under NASA contract NAS5-26555. 
NB gratefully acknowledges financial 
support from the European Research 
Council (ERC-CoG-646928, Multi- Pop). 
and the Royal Society (University 
Research Fellowship). We would also 
like to thank Bill Paxton and the 
MESA community for making this work 
possible. The authors also acknowledge 
Benjamin D. Johnson for helpful insights 
during the development of our methods.

%\bibliography{../master_ref}

\end{document}